\documentclass[sigconf,nonacm]{acmart}
\usepackage{graphicx}
\usepackage[inline]{enumitem}
\usepackage{makecell}
\usepackage{listings}
\usepackage{multirow}
\usepackage{xcolor}
\usepackage{tcolorbox}
\usepackage{color,soul}
\usepackage{tikz}
\usetikzlibrary{positioning,arrows.meta}
\AtBeginDocument{%
  }

\setcopyright{acmlicensed}
\copyrightyear{2026}
\acmYear{2026}
\acmDOI{XXXXXXX.XXXXXXX}
\acmConference[FAccT '26]{The 2026 ACM Conference on Fairness, Accountability, and Transparency}{June 03--05,
  2026}{Montreal, Canada}
\acmISBN{978-1-4503-XXXX-X/2018/06}

\begin{document}

\title{The Stability Trap: Evaluating the Reliability of LLM-Based Instruction Adherence Auditing}

\author{Murtuza N. Shergadwala}
\email{murtuza.shergadwala@workday.com}
\orcid{0000-0003-2337-4809}
\affiliation{%
  \institution{Workday Inc.}
  \city{Pleasanton}
  \state{California}
  \country{USA}
}


\begin{abstract}
 The enterprise governance of Generative AI (GenAI) in regulated sectors, such as Human Resources (HR), demands scalable yet reproducible auditing mechanisms. While ``Large Language Model (LLM)-as-a-Judge'' approaches offer scalability, their reliability in evaluating adherence of different types of system instructions remains unverified. This study asks: \textit{To what extent does the instruction type of an Application Under Test (AUT) influence the stability of judge evaluations?} To address this, we introduce the Scoped Instruction Decomposition Framework to classify AUT instructions into \textit{Objective} and \textit{Subjective} types, isolating the factors that drive judge instability. We applied this framework to two representative HR GenAI applications, evaluating the stability of four judge architectures over variable runs. Our results reveal a ``Stability Trap'' characterized by a divergence between \textit{Verdict Stability} and \textit{Reasoning Stability}. While judges achieved near-perfect verdict agreement ($>99\%$) for both objective and subjective evaluations, their accompanying justification traces diverged significantly. Objective instructions requiring quantitative analysis, such as word counting, exhibited reasoning stability as low as $\approx19\%$, driven by variances in numeric justifications. Similarly, reasoning stability for subjective instructions varied widely ($35\%$--$83\%$) based on evidence granularity, with feature-specific checks failing to reproduce consistent rationale. Conversely, objective instructions focusing on \textit{discrete entity extraction} achieved high reasoning stability ($>90\%$). These findings demonstrate that high verdict stability can mask fragile reasoning. Thus, we suggest that auditors scope automated evaluation protocols strictly: delegate all deterministically verifiable logic to code, while reserving LLM judges for complex semantic evaluation. 
\end{abstract}



\received{13 January 2026}

\maketitle

\section{Introduction}

As Generative AI (GenAI) and Agentic AI systems transition from experimental pilots to critical enterprise infrastructure~\cite{haki2025integrating, yee2024agents}, the mandate for rigorous governance has intensified~\cite{taeihagh2025governance,Judge2024WhenCI,shavit2023practices}. Specifically, in regulated domains such as Human Resources (HR) and Finance, AI-driven decisions regarding talent management or financial reporting carry significant legal and operational weight, demanding uncompromising transparency and accountability~\cite{walkowiak2025generative,oladele2025policy}. Consequently, in such domains, performance alone is insufficient; GenAI systems must be auditable~\cite{taeihagh2025governance,mokander2025blueprint,lam2024framework}.

The sheer scale of generative outputs renders manual auditing untenable, necessitating automated mechanisms~\cite{mokander2022conformity, celestin2019future, adekunle2023integrating}. Consequently, the ``LLM-as-a-Judge'' paradigm~\cite{zheng2023judging} has emerged as a primary candidate, promising a scalable means to verify and monitor models continuously.

An important focus for automated auditing of GenAI systems is at the model layer~\cite{mokander2025blueprint}, specifically the assessment of technical capabilities such as~\textit{instruction adherence}~\cite{young2025models}. Also referred to as \textit{instruction following}~\cite{zhou2023ifeval,ahmed2025specevalevaluatingmodeladherence,qin2024infobench}, this capability denotes the ability of a GenAI Application Under Test (AUT) to strictly comply with the specific constraints and rules defined in its own system prompt. These AUT constraints may range from syntactic requirements, such as formatting outputs as JSON~\cite{RFC8259} or restricting word counts, to nuanced semantic constraints, such as restricting citations to provided context or adhering to specific enterprise tones. Recent literature has rigorously taxonomized such constraints, distinguishing between verifiable objective constraints (such as formatting and word counts) and nuanced subjective dimensions (such as tone and style)~\cite{qin2024infobench,zhou2023ifeval}. These frameworks highlight that verifying a syntactic rule is fundamentally different from evaluating a semantic one.

Yet, the utilization of LLM-as-a-Judge for instruction adherence rests on a critical assumption that the judge can provide a stable, reproducible signal. Recent empirical studies have rigorously quantified the stochastic behavior of LLMs. Notably, Atil et al. \cite{atil2025nondeterminism} demonstrate that even when configured for maximum determinism with model parameters such as temperature, state-of-the-art models exhibit significant non-determinism in their outputs, observing accuracy variations of up to 15\% across identical runs. This introduces a critical reliability crisis~\cite{dobslaw2025challenges} such that if the auditor itself is unstable, the resulting compliance signal becomes noisy, making it difficult to distinguish between a genuine failure of the application and a measurement error by the judge.

While existing literature has extensively benchmarked the general capability of LLMs~\cite{chang2024survey} and quantified their baseline instability as judges~\cite{yu2025ais, li2024llms}, these studies typically evaluate stability in aggregate. They overlook the critical interaction between the nature of the task (as taxonomized by~\cite{qin2024infobench,zhou2023ifeval}) and the reliability of the auditor. Consequently, there remains a critical gap in understanding how the \textit{type of the AUT instruction itself} influences this instability.

Intuitively, one might assume that \textit{Objective} instructions (with a verifiable ground truth) yield more reliable evaluations than \textit{Subjective} ones. We challenge this assumption, arguing that it obscures a deeper risk: that an LLM judge may provide a consistent binary verdict, such as ``Yes'' or ``No'' and ``Pass'' or ``Fail'', based on inconsistent accompanying justifications (hereafter referred to as \textit{reasoning}). To address this gap, this study asks \textbf{RQ}: \textit{To what extent does the instruction type of an AUT influence the stability of instruction adherence evaluations?} We argue that reliability cannot be measured by verdict consistency alone. Using our proposed \textbf{Scoped Instruction Decomposition (SID) Framework}, we disentangle \textit{Verdict Stability} from \textit{Reasoning Stability}. 

Our empirical analysis of two representative HR GenAI use cases reveals a critical ``Stability Trap'': while judges achieved near-perfect binary agreement ($>99\%$) across all instruction types, a stability analysis of their reasoning traces revealed instruction-property-specific divergence. Specifically, objective instructions with quantitative analysis, such as counting, exhibited reasoning stability as low as $\approx19\%$, driven by numeric hallucinations. Such instruction-property-specific differences are important for auditors to consider while automating governance processes for not only GenAI systems but also Agentic AI systems that rely on reasoning traces to adapt, plan, and automate decisions.

We note that this paper does not evaluate any specific [Company] product or production system. Rather, we explore risks in LLM‑based auditing so that enterprise systems, including [Company's], can be designed to avoid those pitfalls.

Our contributions are: (1) We introduce the \textbf{SID Framework} to classify instructions into Objective (Syntactic and Semantic) and Subjective types, isolating the drivers of judge instability; (2) We empirically identify the ``Stability Trap'', demonstrating that high binary agreement often masks underlying reasoning failures, particularly in quantitative syntactic tasks; and (3) We discuss a Hybrid Auditing Protocol for enterprise governance, recommending that all deterministically verifiable logic be delegated to code, while LLMs are reserved for complex semantic evaluation.

\section{Related Work}\label{sec:litrev}

This work intersects AI governance, automated auditing, and the technical assessment of instruction adherence, particularly within regulated industries. High-stakes domains like HR and Finance face stringent mandates for auditability and worker protection~\cite{taeihagh2025governance, mokander2025blueprint, walkowiak2025generative, jiang2025leverage, onwubuariri2024ai, adekunle2023integrating}. To mitigate risks, regulations such as the EU AI Act~\cite{EU_AI_Act_2024} require continuous post-market monitoring~\cite{mokander2022conformity}. While Human-in-the-Loop (HITL) methods offer accuracy, they suffer from scalability limitations~\cite{kumar2024applications, shergadwala2022human}, necessitating automated solutions.

\paragraph{Test Generation Strategies.}
Prior research has operationalized automated auditing through \textit{test case generation}, falling into three categories. \textbf{Standardized benchmarks} like HELM~\cite{liang2022helm}, BIG-bench~\cite{srivastava2023beyond}, DecodingTrust~\cite{wang2023decodingtrust}, and TrustLLM~\cite{huang2024trustllm} provide fixed reference points for capability tracking. \textbf{Collaborative frameworks} leverage human intuition, such as AdaTest~\cite{ribeiro2022adatest}, Dynabench~\cite{kiela2021dynabench}, and Red Teaming at Scale~\cite{ganguli2022redteaming}, to discover failure modes. Finally, \textbf{Perturbation-based auditing} tests robustness via systematic variation, exemplified by CheckList~\cite{ribeiro2020beyond}, AuditLLM~\cite{amirizaniani2024auditllm}, and TextFlint~\cite{wang2021textflint}. To evaluate the resulting volume of test artifacts at scale, the field has increasingly adopted the ``LLM-as-a-Judge'' paradigm~\cite{chang2024survey, li2025generation, liu2023geval, zheng2023judging}. This approach utilizes strong models to approximate human evaluation, becoming the standard mechanism for scoring complex, open-ended generative tasks where simple programmatic assertions are insufficient.

\paragraph{Automated Scoring and Stability.}
Despite it's scalability, the reliability of LLM-as-a-Judge is compromised by the stochastic nature of LLMs~\cite{bender2021dangers}. Auditing contexts face both \textit{adversarial} instability, where judges are manipulated via prompt injection~\cite{li2025llms} and \textit{inherent} instability, or ``Rating Roulette''~\cite{saha2025rating, feuer2025judgment}. \citet{atil2025nondeterminism} demonstrated that significant non-determinism persists even under deterministic configurations. This fragility necessitates scoped evaluation methods that limit their application to tasks where they demonstrate stability.

\paragraph{Instruction Adherence: Verifiable vs. Subjective.}
A critical assessment capability is \textit{instruction adherence}~\cite{young2025models, ahmed2025specevalevaluatingmodeladherence}. Methodologies distinguish between instruction types such as verifiable instructions, tested via programmatic benchmarks like IFEval~\cite{zhou2023ifeval} and INFOBENCH~\cite{qin2024infobench}, and complex subjective instructions~\cite{jiang2024followbench, wen2024benchmarking}. However, relying on LLMs to assess adherence introduces evaluator bias; \citet{zeng2024llmbar} demonstrated with the LLMBar benchmark that judges often prioritize superficial qualities like length and tone over factual constraint satisfaction. To mitigate such inaccuracies, hybrid frameworks like RECAST~\cite{liu2025recast} and IF-Critic~\cite{wen2025if} delegate objective checks to code while reserving LLMs for subjective dimensions. Although these approaches address verdict \textit{accuracy} and bias, existing literature rarely analyzes the \textit{stability} of the judge's underlying reasoning. Specifically, it remains underexplored how instruction verifiability and subjectivity influence the consistency of the judge's justifications, rather than just the correctness of the final score.

\paragraph{The Stability Gap.}
Existing research treats judge instability as noise to be managed via ``LLM Juries''~\cite{verga2024replacing} or calibration~\cite{hashemi2024llm}. However, regulated domains require stability in both verdict and justification, as LLMs often produce ``unfaithful reasoning''~\cite{turpin2023language, lanham2023measuring}. Unlike prior work focusing on verdict correctness~\cite{amirizaniani2024auditllm, li2025llms}, we decouple verdict and reasoning stability to determine if consistent verdicts mask inconsistent reasoning.

\section{Methodology: Scoped Instruction Decomposition}\label{sec:methodology}

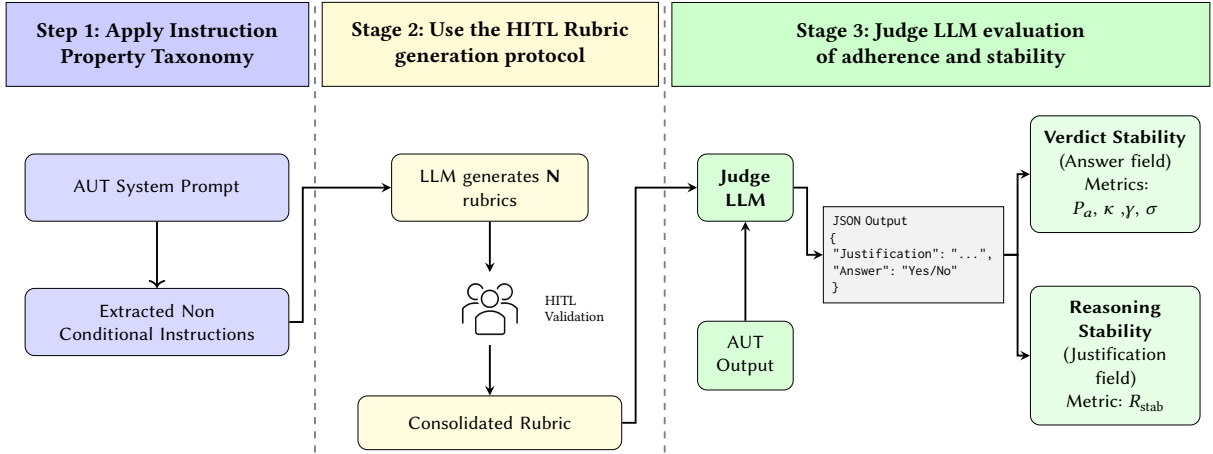
\begin{figure*}[htbp]
\centering
\resizebox{0.9\textwidth}{!}{%
\begin{tikzpicture}[
    font=\sffamily\small,
    box/.style={draw, rounded corners, minimum height=1cm, align=center, inner sep=6pt, fill=white, text width=2.8cm},
    header/.style={draw, rectangle, minimum height=1.2cm, align=center, font=\bfseries, inner sep=8pt, anchor=south west},
    jsonbox/.style={draw, rectangle, fill=gray!10, align=left, font=\ttfamily\scriptsize, text width=2.2cm, inner sep=4pt},
    arrow/.style={->, thick, >=stealth}
]
\def\colwidth{17.5} 
\def\steponeW{4.5} \def\steptwoW{5.0} \def\stepthreeW{8.0} 

\node[header, fill=blue!20, minimum width=\steponeW cm] (h1) at (0, 3.5) {Step 1: Apply Instruction\\Property Taxonomy};
\node[header, fill=yellow!20, minimum width=\steptwoW cm] (h2) at (\steponeW + 0.2, 3.5) {Stage 2: Use the HITL Rubric\\generation protocol};
\node[header, fill=green!20, minimum width=\stepthreeW cm] (h3) at (\steponeW + \steptwoW + 0.4, 3.5) {Stage 3: Judge LLM evaluation\\of adherence and stability};

\draw[dashed, gray, thick] (\steponeW + 0.1, 4.7) -- (\steponeW + 0.1, -2);
\draw[dashed, gray, thick] (\steponeW + \steptwoW + 0.3, 4.7) -- (\steponeW + \steptwoW + 0.3, -2);

\node[box, fill=blue!15, text width=3.5cm] (prompt) at (2.25, 2) {AUT System Prompt};
\node[box, fill=blue!15, text width=3.5cm] (extracted) at (2.25, 0) {Extracted Non\\Conditional Instructions};
\draw[thick, ->] (prompt) -- (extracted);

\node[box, fill=yellow!15, text width=2.5cm] (llm_gen) at (7.2, 2) {LLM generates \textbf{N}\\rubrics};
\node (hitl_img) at (7.2, 0.2) {\includegraphics[width=1.0cm, height=1.0cm, keepaspectratio]{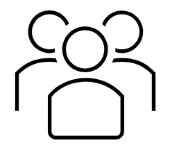}}; 
\node[right=0.1cm of hitl_img, font=\scriptsize, text width=1.5cm] {HITL \\ Validation};
\node[box, fill=yellow!15, text width=3.5cm, minimum height=0.8cm] (rubric) at (7.2, -1.5) {Consolidated Rubric};

\node[box, fill=green!15, text width=1cm] (judge) at (11.0, 2) {\textbf{Judge LLM}};
\node[box, fill=green!15, text width=1cm] (output) at (11.0, -0.5) {AUT Output};
\node[jsonbox, text width=2.4cm] (json_artifact) at (13.5, 1) {JSON Output \\ \{ \\ "Justification": "...",\\ "Answer": "Yes/No"\\ \}};
\node[box, fill=green!10, text width=2.1cm, minimum height=1.5cm] (verdict) at (16.5, 2.2) {\textbf{Verdict Stability}\\(Answer field)\\Metrics: $P_a$, $\kappa$ ,$\gamma$, $\sigma$};
\node[box, fill=green!10, text width=2.1cm, minimum height=1.5cm] (reasoning) at (16.5, -0.5) {\textbf{Reasoning Stability}\\(Justification field)\\Metric: $R_{\mathrm{stab}}$};

\draw[arrow] (extracted.east) -- ++(0.2,0) |- (llm_gen.west);
\draw[arrow] (llm_gen) -- (hitl_img);
\draw[arrow] (hitl_img) -- (rubric);
\draw[arrow] (rubric.east) -- ++(0.2,0) |- (judge.west);
\draw[arrow] (output) -- (judge);
\draw[arrow] (judge.east) -- ++(0.2,0) |-(json_artifact.west);
\draw[arrow] (json_artifact.east) -- ++(0.2,0) |- (verdict.west);
\draw[arrow] (json_artifact.east) -- ++(0.2,0) |- (reasoning.west);
\end{tikzpicture}%
}
\caption{The Scoped Instruction Decomposition (SID) Framework. Step 1 extracts instructions; Stage 2 validates them via HITL; Stage 3 evaluates Verdict and Reasoning stability independently.}
\Description[SID Framework Flowchart]{Flowchart of the SID framework showing three stages: Instruction Taxonomy, HITL Rubric Generation, and Judge LLM Evaluation.}
\label{fig:framework_pipeline}
\end{figure*}

To answer \textbf{RQ}, we introduce the \textit{Scoped Instruction Decomposition} (SID) Framework (Figure~\ref{fig:framework_pipeline}). It operates in three stages that include: (1) applying an Instruction Property Taxonomy to classify and scope the prompt instructions of any GenAI AUT; (2) transforming these scoped instructions into a deterministic, binary rubric via a Human-in-the-Loop (HITL) Protocol; and (3) evaluating the AUT's output against this rubric using a separate Judge LLM. 


Designed to operationalize the auditing of adherence, this framework builds upon the ``Decomposed Requirements Following Ratio'' (DRFR)~\cite{qin2024infobench}. However, we extend DRFR via a ``scoping'' mechanism, arguing that not all decomposed instructions are valid candidates for automated testing. We systematically filter instructions based on properties that influence judge reliability.

\subsection{The Instruction Property Taxonomy}\label{subsec:taxonomy}
\begin{figure}[htbp]
    \centering
    \resizebox{0.75\columnwidth}{!}{
    \begin{tikzpicture}[
        font=\small\sffamily,
        box/.style={draw, rectangle, minimum width=3.5cm, minimum height=2cm, align=center, text width=3.1cm},
        label_text/.style={font=\bfseries\sffamily}
    ]
    \node[box, fill=green!10] (q1) at (0, 0) {\textbf{Quadrant I}\\Objective \&\\ Non-conditional\\\scriptsize \textit{(e.g., "Output in JSON format")}};
    \node[box, fill=yellow!10] (q2) at (3.7, 0) {\textbf{Quadrant II}\\Subjective \&\\ Non-conditional\\\scriptsize \textit{(e.g., "Maintain a supportive tone")}};
    \node[box, fill=blue!10] (q3) at (0, -2.5) {\textbf{Quadrant III}\\Objective \&\\ Conditional\\\scriptsize \textit{(e.g., "If eligible, add sentence X")}};
    \node[box, fill=red!10] (q4) at (3.7, -2.5) {\textbf{Quadrant IV}\\Subjective \&\\ Conditional\\\scriptsize \textit{(e.g., "If angry, apologize")}};
    
    \node[above=0.1cm of q1, label_text] {\small Objective};
    \node[above=0.1cm of q2, label_text] {\small Subjective};
    \node[rotate=90, label_text] at (-2.1, 0) {\small Non-Conditional};
    \node[rotate=90, label_text] at (-2.1, -3.0) {\small Conditional};
    
    \draw[->, thick] (-2.4, 2.0) -- (5.5, 2.0) node[midway, above] {\textbf{Verifiability}};
    \draw[->, thick] (-2.4, 2.0) -- (-2.4, -4) node[midway, above, rotate=90] {\textbf{Dependency}};
    \end{tikzpicture}%
    }
    \caption{Instruction Classification Taxonomy.}
    \Description[A 2x2 classification matrix categorizing prompt instructions along two axes: Verifiability (Objective vs. Subjective) and Dependency (Non-Conditional vs. Conditional), resulting in four distinct quadrants of instruction types.]{The figure presents a 2x2 matrix titled "Classification Taxonomy of Prompt Instructions." The horizontal axis represents "Verifiability," divided into "Objective" on the left and "Subjective" on the right. The vertical axis represents "Dependency," divided into "Non-Conditional" at the top and "Conditional" at the bottom.

This creates four quadrants:

Quadrant I (Top-Left): Represents "Objective & Non-conditional" instructions. The provided example is: "Output in JSON format."

Quadrant II (Top-Right): Represents "Subjective & Non-conditional" instructions. The provided example is: "Maintain a supportive tone."

Quadrant III (Bottom-Left): Represents "Objective & Conditional" instructions. The provided example is: "If promotion_eligibility is true, add sentence X."

Quadrant IV (Bottom-Right): Represents "Subjective & Conditional" instructions. The provided example is: "If the user seems angry, begin with a polite apology."}
    \label{fig:taxonomy_matrix}
\end{figure}
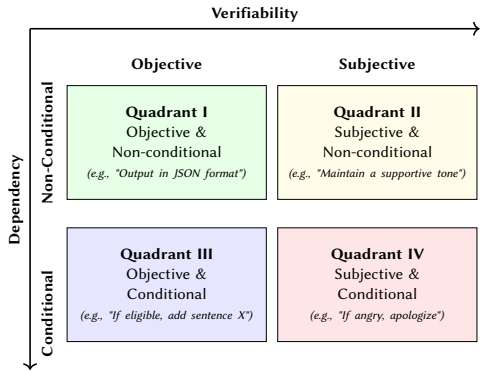

We classify instructions along two orthogonal dimensions: Verifiability and Dependency (Figure~\ref{fig:taxonomy_matrix}).

\paragraph{Dimension 1: Verifiability.} We classify instructions based on the existence of a ground truth. \textbf{Objective Instructions ($I_{obj}$)} allow for strictly True/False verification. We further distinguish between \textit{Syntactic Objectivity} (formatting, word counts)~\cite{zhou2023ifeval} and \textit{Semantic Objectivity} (factual adherence requiring reading comprehension). Conversely, \textbf{Subjective Instructions ($I_{subj}$)} rely on qualitative interpretation of tone or style. These lack a universal ground truth, making them vulnerable to interpretation variance.

\paragraph{Dimension 2: Dependency.} We separate \textbf{Non-Conditional Instructions (NCI)}, which apply universally (e.g., ``Always sign off with X''), from \textbf{Conditional Instructions (CI)}, which are logic-gated (e.g., ``If eligible, do X'').

\paragraph{Operational Scoping and Counterfactual Blindness.} We restrict evaluation to \textbf{Non-Conditional Instructions} (Quadrants I and II). We exclude conditionals due to a theoretical \textbf{``Counterfactual Blindness''} in reference-free auditing: if a conditional rule ``If X, do Y'' results in a missing Y, a judge cannot distinguish between a ``Correct Non-Action'' (X was false) and a ``Failure'' (Model ignored Y) without external state verification. Testing this requires counterfactual input manipulation, which introduces variables extraneous to the intrinsic stability of the judge.

\subsection{Rubric Generation and Judge Design}\label{subsec:hitl}
\paragraph{HITL Protocol.} We employ a Human-in-the-Loop protocol to generate the test rubric. First, an LLM is prompted to act as a Quality Assurance (QA) Engineer to parse the prompt, extracting candidate Yes/No questions (see Appendix~\ref{appendix:rubricGenObj}-- ~\ref{appendix:rubricGenSubj}). Crucially, we exclude ambiguous scales (e.g., 1-5 ratings). Second, Subject Matter Experts (SMEs) validate these candidates. The SME's role is not just review, but to \textit{verify the existence of a binary ground truth} for objective items and ensure no conditional logic has leaked into the test set.

\paragraph{Judge Configuration.} We employ a reference-free Judge LLM. The model returns a structured JSON object containing: (1) \texttt{Question:} for traceability; (2) \texttt{Justification:} a natural language rationale; and (3) \texttt{Answer:} the binary verdict. This structure is essential to decouple the verdict from the reasoning trace (see Appendix~\ref{appendix:judgePrompt}).

\subsection{Reliability Metrics}\label{subsec:metrics}
We quantify stability across $k$ repeated independent runs of the Judge.

\subsubsection{Verdict Stability Metrics}

\paragraph{Primary Metric: Percentage Agreement ($P_a$)}
At the question level, we utilize raw \textit{Percentage Agreement} as our primary metric for verdict stability. This is calculated as the proportion of repeated runs that yield the exact same  \texttt{Answer} for a specific rubric question. This choice is necessitated to analyze scenarios where a judge exhibits perfect determinism (such as selecting ``Yes'' in 20 out of 20 runs), the variance of the ratings is zero. In these cases of unanimous agreement, chance-corrected metrics, such as Fleiss' Kappa discussed below, can become undefined or unstable. 

\paragraph{Chance-Corrected Agreement ($\gamma$ and $\kappa$)}
To distinguish genuine verdict stability from random chance, we employ \textit{Gwet's AC1} ($\gamma$)~\cite{gwet2008computing} as our primary statistic, while reporting \textit{Fleiss' Kappa} ($\kappa$) for reference.

We prioritize AC1 due to the well-documented ``Kappa Paradox''~\cite{feinstein1990high, wongpakaran2013comparison}: in scenarios where a judge exhibits high verdict consensus (such as $>95\%$ agreement), Kappa often yields paradoxically lower scores ($<0.8$) due to skewed marginal probabilities. Gwet's AC1 assumes that chance agreement occurs only when raters are uncertain, making it a more robust estimator for auditing tasks where near-perfect agreement is expected. We interpret AC1 values using standard Landis and Koch benchmarks~\cite{landis1977measurement} (such as $\gamma > 0.81$ indicates ``Almost Perfect'' reliability).

\paragraph{Aggregate Metric: Volatility of Adherence Score ($\sigma$)}
At the rubric level, we assess the verdict stability of the overall instruction adherence assessment. We define the \textit{Instruction Adherence Score} ($S_i$) for a single run $i$ as the normalized count of compliant features:
\begin{equation}
S_i = \frac{\sum \text{Yes Responses}}{\text{Total Questions}}
\end{equation}
To quantify stability, we calculate the Standard Deviation ($\sigma$) of $S$ across all $k$ runs. Crucially, this distinguishes reliability from performance; a low $\sigma$ implies that the auditor provides a reproducible adherence signal, whereas a high $\sigma$ indicates that the scoring of the judge fluctuates across runs.

\subsubsection{Reasoning Stability Metric ($R_{\mathrm{stab}}$)}\label{subsubsec:reas_met}
While verdict-based metrics measure the consistency of the final output, they fail to capture the validity of the derivation. To address this, we introduce \textit{Reasoning Stability}, a metric quantified via a two-stage clustering pipeline applied to the Judge's natural language \texttt{Justification} field.

\textit{Stage 1: Heuristic Fingerprinting.}
First, we apply strategy-based extraction to isolate the core rationale from the justification. Based on the instruction type, we employ targeted regular expressions. \textbf{\textit{Numeric Strategy (Syntactic):}} For \textit{quantitative constraints} (such as word limits), we extract the primary verifiable integer value (such as parsing ``155'' from ``The text contains 155 words''), stripping punctuation and whitespace as well as filtering out unconnected numerals. \textbf{\textit{Quote Strategy (Semantic):}} For \textit{verbatim constraints} (such as specific phrase inclusion), we extract substrings enclosed in quotation marks. \textbf{\textit{Assertion Strategy (Subjective):}} For \textit{qualitative constraints} (such as tone or style), we normalize the assertion text by removing stop words and punctuation to cluster semantically equivalent prose.

\textit{Stage 2: Semantic Clustering.}
To account for linguistic variance, we embed the extracted fingerprints using a sentence transformer model (\texttt{all-MiniLM-L6-v2})\footnote{\url{https://huggingface.co/sentence-transformers/all-MiniLM-L6-v2}} from the \texttt{sentence-transformers} library~\cite{reimers-2019-sentence-bert}. This model is a distilled version of BERT optimized for semantic similarity tasks~\cite{wang-2020-minilm}. Then we cluster the fingerprints using Density-Based Spatial Clustering of Applications with Noise (DBSCAN)~\cite{schubert2017dbscan}.

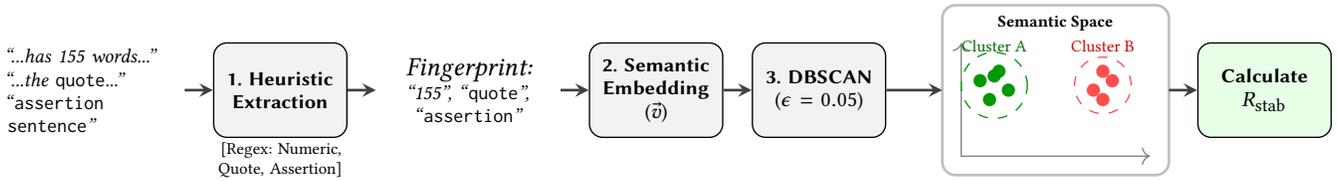
\begin{figure*}[ht]
    \centering
    \resizebox{\textwidth}{!}{%
    \begin{tikzpicture}[
        font=\sffamily\scriptsize,
        node distance=0.4cm and 0.3cm, 
        process/.style={
            rectangle, 
            draw, 
            rounded corners, 
            fill=gray!10,
            text width=1.2cm, 
            minimum height=1.0cm, 
            align=center,
            inner sep=3pt,
            anchor=center 
        },
        data/.style={
            text width=1.8cm, 
            align=center, 
            font=\itshape\scriptsize,
            anchor=center,
            inner sep=2pt
        },
        arrow/.style={->, thick, >=stealth, gray!50!black}
    ]

    \node[data, align=left] (input) {
        ``...has 155 words...'' \\ 
        ``...the \texttt{quote}...''\\
        ``\texttt{assertion sentence}''
    };

    \node[process, right=of input] (step1) {\textbf{1. Heuristic Extraction}};
    
    \node[anchor=north, font=\tiny, align=center, text width=2.5cm, inner sep=1pt] 
        (regex) at (step1.south) {
        [Regex: Numeric, Quote, Assertion]
    };

    \node[data, right=of step1] (fingerprint) {
        {\small Fingerprint:} \\
        ``155'', ``\texttt{quote}'', ``\texttt{assertion}''
    };

    \node[process, right=of fingerprint] (step2) {\textbf{2. Semantic Embedding} \\ ($\vec{v}$)};

    \node[process, right=of step2] (step3) {\textbf{3. DBSCAN} \\ ($\epsilon=0.05$)};

    \begin{scope}[shift={(9, 0)}, local bounding box=plotbox] 
        
        \draw[thick, gray!50, rounded corners] (0, -0.9) rectangle (2.4, 0.9);
        
        \node[font=\bfseries\tiny, anchor=north] at (1.2, 0.9) {Semantic Space};
        
        \draw[<->, gray] (0.2, 0.5) -- (0.2, -0.7) -- (2.2, -0.7);

        \foreach \x/\y in {0.4/0.1, 0.5/-0.1, 0.6/0.2, 0.7/0.0, 0.55/0.15}
            \fill[green!60!black] (\x, \y) circle (2pt);
        \node[font=\tiny, green!40!black, anchor=south] at (0.55, 0.3) {Cluster A};
        \draw[dashed, green!60!black] (0.55, 0.05) circle (0.35cm);

        \foreach \x/\y in {1.6/0.0, 1.7/0.2, 1.8/0.1, 1.7/-0.1}
            \fill[red!70] (\x, \y) circle (2pt);
        \node[font=\tiny, red!80!black, anchor=south] at (1.7, 0.3) {Cluster B};
        \draw[dashed, red!70] (1.7, 0.05) circle (0.3cm);
        
    \end{scope}

    \node[process, fill=green!10, right=0.3cm of plotbox] (calc) {\textbf{Calculate} \\ $R_{\mathrm{stab}}$};

    \draw[arrow] (input) -- (step1);
    \draw[arrow] (step1) -- (fingerprint);
    \draw[arrow] (fingerprint) -- (step2);
    \draw[arrow] (step2) -- (step3);
    
    \draw[arrow] (step3) -- (plotbox.west);
    \draw[arrow] (plotbox.east) -- (calc);

    \end{tikzpicture}
    }
    \caption{The Reasoning Stability Pipeline. Raw justification strings are distilled into fingerprints, embedded, and clustered. The schematic scatter plot illustrates the separation of dominant reasoning traces (Cluster A) from variants (Cluster B).}
    \Description[Pipeline diagram for Reasoning Stability]{This figure illustrates the multi-stage pipeline for calculating Reasoning Stability ($R_{stab}$). The left column displays the sequential data processing flow: raw input justification text snippets (top) undergo "Heuristic Extraction" using specific regex strategies (Numeric, Quote, Assertion) to create a standardized "Fingerprint." This fingerprint is then converted into a high-dimensional "Vector $v$" via "Semantic Embedding." These vectors are subsequently grouped using the "DBSCAN" clustering algorithm with $\epsilon=0.05$. The right side of the figure features a schematic 2D scatter plot visualizing this high-dimensional semantic space, where a tightly grouped "Dominant" cluster (Cluster A, green dots) is clearly separated from a "Variant" cluster (Cluster B, red dots). The pipeline concludes with the final step to "Calculate $R_{stab}$" based on these cluster assignments.}
    \label{fig:reasoning_pipeline_horiz}
\end{figure*}

To ensure rigorous stability measurement, we configured DBSCAN with a strict similarity threshold ($\epsilon=0.05$) and cosine distance metric. We empirically calibrated this hyperparameter to ensure a unified clustering pipeline: specifically, this strict threshold is required to differentiate distinct integer values (such as distinguishing ``155'' from ``156'') which otherwise exhibit high proximity in semantic vector space. This configuration groups only semantically equivalent rationales while correctly treating slight deviations in logic as distinct reasoning traces.

We define Reasoning Stability ($R_{\mathrm{stab}}$) for a given question as the size of the dominant semantic cluster ($|C_{\mathrm{dominant}}|$) relative to the total number of runs ($N$): $R_{\mathrm{stab}} = (|C_{\mathrm{dominant}}| / N) \times 100$. A low $R_{\mathrm{stab}}$ indicates that the judge is hallucinating different evidence (such as different word counts) or shifting its rationale across runs.

\section{Experimental Setup}\label{sec:experiment}

In this section, we discuss our experimental setup. We first describe the HR GenAI use cases and the application of the \textit{Scoped Instruction Decomposition} (SID) framework (introduced in Section~\ref{sec:methodology}). We then detail the experimental data generation pipeline and model configurations used for both the AUT and the judges. Finally, we discuss the operationalization of stability analysis for the experimental setup.

\subsection{Use Case Selection: Enterprise HR GenAI Features} We selected two distinct enterprise HR GenAI application scenarios to simulate real-world business constraints. 

\paragraph{AUT Scenario A: HR Performance Review}
This scenario simulated an HR task where the model acts as a supporting tool for managers. The objective was to synthesize raw employee feedback data into a formal performance review. 
\begin{itemize}
    \item \textit{Input Variables:} Employee Name, Review Period, Key Strengths, Areas for Growth, Project Examples, and Promotion Readiness.
    \item \textit{Key Constraints:} The prompt enforced strict syntactic rules, such as formatting the output as a JSON object with specific keys and requiring verbatim inclusion of specific project examples. It also included semantic constraints, such as using active voice and avoiding absolute terms like ``always'' or ``never.'' Refer to Appendix~\ref{appendix:appPromptA} for details.
\end{itemize}

\paragraph{AUT Scenario B: Career Vision Statement}
This scenario simulated a career coaching task. The objective was to synthesize a worker's professional history into a compelling ``Career Vision Statement.''
\begin{itemize}
    \item \textit{Input Variables:} Worker Name, Current Role, Work History, Career Goals, Top Skills, Skill Interests, Development Items, Job Interests, and Recent Feedback.

    \item \textit{Key Constraints:} Instructions included writing in a ``highly motivational and aspirational'' tone, crafting a ``punchy, memorable personal brand slogan,'' and demonstrating empathy. Objective constraints included a 500-word limit and the mandatory inclusion of specific skills verbatim. Refer to Appendix~\ref{appendix:appPromptB} for the full prompt.
\end{itemize}
\textbf{These were chosen based on the risk to a worker's long-term economic opportunities}, such as promotions (Scenario A) and career trajectories (Scenario B), ensuring our stability findings generalize across different AUT instruction types.
\subsection{Application of the SID Framework}
For our stability analysis, we applied the SID framework to the system prompts of both scenarios.

\subsubsection{Rubric Construction Protocol}
We utilized the HITL methodology described in Section~\ref{subsec:hitl}. A high-reasoning LLM (Gemini 3 Pro~\cite{gemini3pro2025}) was used initially to parse the system prompts and generate candidate atomic questions. These drafts were then manually reviewed by SMEs to ensure measurability before being classified according to the Instruction Property Taxonomy.

\subsubsection{The Resulting Test Sets}
The application of the taxonomy yielded two distinct test sets for each scenario, effectively isolating the variable of instruction verifiability. The complete rubrics are detailed in Tables 1-4 (refer to Appendix~\ref{appendix:evalrubrics}).

\textbf{Scenario A (HR) Test Set:}
\begin{itemize}
    \item \textbf{Objective Set ($I_{obj}$):} Focuses on verifiable adherence. Criteria include: ``Is the output a valid JSON object?'', ``Does the output contain at $\geq2$ exact string matches from \{project\_examples\}?'', and ``Is the word count $\leq$ 200?''.
    \item \textbf{Subjective Set ($I_{subj}$):} Focuses on tonal adherence. Criteria include: ``Does the output have a constructive tone?'', ``Is the feedback clear and direct?'', and ``Does the style encourage future improvement?''.
\end{itemize}

\textbf{Scenario B (Career Vision Statement) Test Set:}
\begin{itemize}
    \item \textbf{Objective Set ($I_{obj}$):} Focuses on structural integrity. Criteria include: ``Do both text values contain at least one first-person pronoun?'', ``Does the headline begin strictly with a present participle verb?'', and ``Does it contain at least two skills found in the \{top\_skills\} list?''.
    \item \textbf{Subjective Set ($I_{subj}$):} Focuses on creative quality. Criteria include: ``Is the overall tone highly motivational?'', ``Does the response demonstrate empathy?'', and ``Does the vision statement feel action-oriented?''.
\end{itemize}

\subsection{Experimental Data and Model Configurations}

\subsubsection{Data Privacy and Synthetic Data}
Given the highly sensitive nature of Human Resources and Talent Management data, utilizing proprietary customer data for this study was precluded by strict confidentiality agreements, regulations, and organizational policies regarding Personally Identifiable Information (PII). To address this while ensuring experimental validity, we employed a \textit{structurally faithful} synthetic data generation approach.

While the content is synthetic, the \textit{schema design and input feature sets} were designed to mirror the functional requirements of real-world production use cases. This approach offers two distinct advantages:
\begin{enumerate*}
    \item \textit{Ethical Compliance:} It creates a firewall against data leakage, ensuring no actual worker data is exposed or processed during the experimental validation.
    \item \textit{Feature Parity:} By generating inputs based on actual production feature definitions (such as using the exact JSON keys and variable types found in the live system), we ensure the judges are evaluated against the same structural complexity encountered in the wild, preserving the validity of the stability signal.
\end{enumerate*}

\subsubsection{Input Data for AUTs}
To generate the input variables for both scenarios, we utilized Google Gemini 3 Pro~\cite{gemini3pro2025}. This model was selected for its high reasoning capability to generate diverse and coherent user profiles but was excluded from the testing pool as it was not used by any AUT in production. We generated $80$ unique profiles across Scenario A ($50$) and B ($30$). These profiles were prompted to simulate diverse production scenarios defined by SMEs, such as varying tenure lengths, diverse skill taxonomies, and mixed performance feedback sentiments, ensuring the AUT was tested against a realistic variance of inputs.

\subsubsection{AUT Output Generation (The Corpus)}
The four models under test (defined below) acted as the GenAI AUT, generating a response for every unique input profile across both scenarios. This resulted in a total corpus of $320$ unique outputs ($80$ total inputs $\times$ $4$ models). To maximize stability, all outputs were generated with \texttt{temperature=0}.

\subsubsection{Model Configurations}
We selected four state-of-the-art LLMs to serve as the core computational engines. These models were utilized in a dual capacity: first as \textit{Generative Agents} to produce the corpora, and subsequently as \textit{Automated Evaluators} (Judges) to score each corpus. We employed a full-factorial cross-evaluation design, where every judge model evaluated the outputs of every generator model.

\paragraph{Model Selection Strategy:}
We deliberately selected a mix of \textit{Proprietary} and \textit{Open-Weights} models, spanning a parameter variance from ~8B to 405B. This diversity allows us to verify if stability is a function of model scale or training methodology, rather than being specific to a single architecture. The following models were chosen as they were available under our specific organizational agreements for use.
\begin{itemize} 
    \item \textit{Proprietary Models:} Anthropic Claude 3.5 Sonnet~\cite{claude35sonnet2024} and Google Gemini 2.5 Flash Lite~\cite{gemini25flashlite2025}. 
    \item \textit{Open-Weights Models:} Meta Llama 3.1 405B~\cite{llama31modelcard2024} and DeepSeek R1 Distill Llama 8B~\cite{deepseekr12025}. 
\end{itemize}

\paragraph{Judge Configuration:}
Crucially, when acting as Judges, all models were configured with \texttt{temperature=0}. While recent literature on reasoning-intensive models, such as DeepSeek-R1, suggests that non-zero temperatures (e.g., $0.6$) may facilitate optimal chain-of-thought exploration and prevent repetition loops~\cite{deepseekr12025}, we deliberately selected greedy decoding to simulate a strict enterprise auditing environment where reproducibility is paramount.  This configuration is intended to maximize inherent determinism, ensuring that any observed instability in the scoring is driven by the model's inability to resolve the prompt consistently rather than by sampling randomness.

\subsection{Stability Analysis Protocol}
To validate the consistency of the LLM-as-a-Judge methodology, we executed a repeated-measures design. Each of the four judge models evaluated every unique observation in the corpus across multiple independent experimental groups.

We established five experimental groups defined by the number of repeated runs $k \in \{5, 10, 20, 25, 30\}$. Cumulatively, this resulted in $90$ distinct evaluation runs per input. Given the corpora totaling $320$ outputs ($80$ combined inputs $\times$ 4 AUTs), this protocol generated $28,800$ judgments per judge ($320\times90$) and a total of $115,200$ experimental judgments across four architectures. To synthesize this raw data into the reported results, we applied the aggregation protocols corresponding to the reliability metrics defined in Section~\ref{subsec:metrics}.

\subsubsection{Verdict Stability ($P_a, \kappa, \gamma$) and Rubric-Level Stability ($\sigma$) Aggregation}\label{subsubsec:agg_strat}
Given the factorial design (4 AUT Generators $\times$ 4 Judges), pooling raw verdict data would conflate inter-generator variability with judge instability. To rigorously isolate the systemic performance of the methodology, we employed a \textit{Stratified Aggregation} protocol:
\begin{enumerate*}[label=(\arabic*)]
    \item \textit{Pair-Wise Calculation:} We first calculated the stability metrics ($P_a$, $\kappa$, $\gamma$) at the question level across runs for every input profile, for each distinct Judge-Generator pair ($4 \times 4 = 16$ pairs). These question-level scores were then averaged at the rubric level to derive a single aggregate stability score. 
    \item \textit{Global Macro-Averaging:} The final values reported in Section~\ref{sec:results} represent the macro-average across \textit{all 16 pairs}. This aggregates performance across both the generator source and the judge architecture, effectively providing a global stability index for the ``LLM-as-a-Judge'' paradigm itself, rather than evaluating specific model capabilities.
\end{enumerate*} For the instruction adherence score stability ($\sigma$), we followed the same stratified approach. We computed the standard deviation $\sigma$ of the \textit{Instruction Adherence Score} across runs per profile per scenario for each Judge-Generator pair first. These values were then averaged across all the $16$ pairs to report the expected volatility of a standard automated audit.

\subsubsection{Reasoning Stability Operationalization ($R_{stab}$)}
The operationalization of Reasoning Stability required a localized clustering approach to account for test case (profile) independence. (1) \textit{Scope of Clustering:} We performed semantic clustering for every unique combination of Input Artifact, AUT Generator, Judge, Scenario, Objective/Subjective Test Set, and Question collecting the set of justifications across $k$ runs. (2) \textit{Isolation of Reasoning Traces:} This localized scope ensures we measure the internal consistency of evaluating a \textit{single specific artifact} (e.g., ``Did the judge A consistently cite `155 words' for summary X of Profile 1 by AUT B?''), isolating self-consistency from input variation. (3) \textit{Aggregation:} The $R_{stab}$ score was calculated for each such observation and then averaged using the global stratified protocol described above to produce the final macro score.

\section{Results}\label{sec:results}

Our experimental analysis reveals a stark dichotomy between the stability of the binary verdict and the stability of the underlying reasoning. In this section, we first present the high consensus observed in verdict stability (Section~\ref{subsec:verdict_results}), followed by the significant divergence uncovered in reasoning stability (Section~\ref{subsec:reasoning_results}).

\subsection{Verdict Stability: High Agreement and the Kappa Paradox}\label{subsec:verdict_results}
We treated each Judge architecture as an independent auditor. For every question-instruction pair, we calculated stability metrics (Percentage Agreement $P_a$, Gwet’s AC1 $\gamma$, Fleiss’ Kappa $\kappa$) individually for each model across all its experimental runs ($k=90$). Crucially, in instances where perfect unanimity ($100\%$ agreement) rendered Fleiss' Kappa mathematically undefined, we imputed a value of $\kappa=1.0$ to accurately reflect the observed total consensus in our aggregate reporting.

The results presented in Table~\ref{tab:verdict_summary_main} represent the macro-average of these model-specific scores (refer to Section~\ref{subsubsec:agg_strat} for the aggregation approach.) Detailed question-level breakdowns are available in Appendix~\ref{appendix:verdict_detailed_results}. As observed, the average verdict stability was exceptionally high, with $P_a$ exceeding $99\%$ across all scenarios. This indicates that the chosen LLMs, when acting as judges with \texttt{temperature=0}, exhibit high verdict stability for binary adherence evaluation tasks.

\begin{table*}[htbp]
\centering
\small
\renewcommand{\arraystretch}{1.25}
\setlength{\tabcolsep}{6pt}

\caption{\textbf{Verdict Stability Summary (Aggregate)}. 
Values represent the macro-average of stability statistics calculated independently for each of the $N=16$ judge generator pairs across all questions within the specific scenario and instruction type ($N=90$ pooled runs).}
\label{tab:verdict_summary_main}

\begin{tabular}{l l c c c | c}
\toprule
\textbf{Scenario} & \textbf{Type} & \textbf{Avg. Agree (\%)} & \textbf{Avg. AC1 ($\gamma$)} & \textbf{Avg. Kappa ($\kappa$)} & \textbf{Rubric $\sigma$ (Avg [Range])} \\
\midrule
\multirow{2}{*}{\textbf{A: HR Review}} 
 & Objective & $99.88$ & $0.9976$ & $0.9731$ & $0.3158~[0.0000 - 1.3725]$ \\
 & Subjective & $99.98$ & $0.9997$ & $0.9723$ & $0.0304~[0.0000-0.1287]$ \\
\midrule
\multirow{2}{*}{\textbf{B: Career Vision Statement}} 
 & Objective & $99.30$ & $0.9860$ & $0.7919$ & $1.4176~[0.0439-3.4881]$ \\
 & Subjective & $99.99$ & $0.9998$ & $0.9979$ & $0.0290~[0.0000-0.1363]$ \\
\bottomrule
\end{tabular}
\end{table*}

\paragraph{Validating the Kappa Paradox}
Our results empirically validate the necessity of using Gwet's AC1 over Fleiss' Kappa for compliance auditing. Despite imputing $\kappa=1.0$ for perfect cases, the Kappa scores remained highly volatile due to the ``Kappa Paradox''~\cite{feinstein1990high, wongpakaran2013comparison} in scenarios with high pass rates. A relevant example is observed in the Scenario B Objective $Q_2$ (Headline Active Verb Check). As detailed in Table~\ref{tab:scenario_b_objective_verdict} (Appendix~\ref{appendix:verdict_detailed_results}), the judges achieved a pooled agreement of \textbf{$99.80\%$}, indicating near-universal consensus. However, the corresponding Fleiss' Kappa dropped to \textbf{0.523}, a value traditionally interpreted as merely ``Moderate'' agreement. In contrast, Gwet's AC1 yielded a score of \textbf{0.996}, correctly aligning with the observed stability. This discrepancy confirms that low Kappa values in this domain are statistical artifacts of skewed distributions rather than evidence of actual judge disagreement.

\paragraph{The Subjective Stability and Score Volatility}
Counterintuitively, Subjective instructions exhibited higher verdict stability than Objective ones. As shown in Table~\ref{tab:verdict_summary_main}, Subjective criteria in Scenario B achieved near-perfect \textbf{99.99\% agreement} and \textbf{0.9998 AC1}. This resulted in effectively static aggregate scores with negligible volatility ($\sigma \approx 0.03$), indicating that the final pass/fail verdicts for qualitative criteria remained virtually identical across all 90 runs.

In contrast, the Objective rubric in Scenario B exhibited a standard deviation of $\sigma = 1.42$, nearly $50$ times higher than their subjective counterparts. While this absolute volatility is low, it highlights a distinct behavioral difference. In this experimental setting, the subjective evaluations remained effectively static across all runs, whereas the objective evaluations exhibited greater fluctuations. This indicates that strictly verifiable tasks were paradoxically more prone to scoring inconsistency than qualitative ones when using LLMs as judges.

\noindent\textbf{Impact of Model Size.} Stratifying our analysis by judge LLM parameter size reveals that while large models (400B+) achieve near-perfect agreement, small models (8B) occasionally falter on \textit{Compound Constraints}. While generally stable ($>98\%$ agreement), small model agreement dropped to $88.6\%$ on specific compound constraints (e.g., Scenario A's $Q_1$), suggesting that instruction complexity can strain the attention mechanisms of smaller architectures.

\paragraph{The Stability Trap}
If an auditor were to rely solely on these verdict-based metrics, the conclusion would be that LLM judges are highly robust and production-ready. However, the near-universal consensus, especially on subjective tasks, raises a critical question: \textit{Is the judge actually evaluating the criteria, or simply hallucinating a justification to support a default `Yes' verdict?} This necessitates the analysis of Reasoning Stability.

\subsection{Reasoning Stability}\label{subsec:reasoning_results}

Mirroring our approach for verdict stability, we calculated the \textit{Reasoning Stability Score} ($R_{stab}$) for each judge architecture independently before deriving the macro-averages presented in this section. This ensures that our metric captures \textit{intra-model} semantic consistency, effectively measuring whether a specific judge architecture consistently cites the same evidence for its decisions across repeated trials, rather than conflating reasoning styles across different architectures.

Unlike verdict stability, which remained consistently high ($>99\%$) across all conditions, reasoning stability was highly variable. Our analysis indicates this variance is not random, but is driven by the interaction between, \begin{enumerate*}
    \item the \textbf{nature} of the adherence task (as characterized by our SID Taxonomy),
    \item the \textbf{breadth of the evidence space} available for citation,
    \item  and the resulting \textbf{justification strategy} (Numeric, Quote, or Assertion).
\end{enumerate*}

Specifically, we observed that the stability of both \textit{Objective} and \textit{Subjective} instructions was governed by the \textit{nature} of these factors. Regarding the \textit{breadth of the evidence space}, tasks requiring extraction from a fuzzy or continuous search space, whether estimating continuous word counts (Objective) or isolating ambiguous terms (Subjective), yielded unstable rationales ($\approx 19\%-35\%$), as judges did not converge on a single evidence target. Furthermore, regarding \textit{justification strategy}, we found that tasks allowing for multiple reasoning paths, such as choosing between an \textit{Assertion} vs. a \textit{Quote}, induced strategy oscillation, further degrading reproducibility. Conversely, tasks with an evidence target in the input (e.g., discrete entity extraction), constrained the judges to a single strategy and remained robust ($>90\%$).

\begin{table*}[htbp!]
\centering
\small
\renewcommand{\arraystretch}{1.2}
\setlength{\tabcolsep}{6pt} 
\caption{\textbf{Objective Reasoning Stability (Both Scenarios)}. Sorted by Pooled Stability. \textbf{Scenario} A=HR Review, B=Career Vision. \textbf{Strat.} denotes the rationale extraction strategy from justifications: (N)umeric, (Q)uote, (A)ssertion.}
\label{tab:objective_reasoning_consolidated}
\begin{tabular}{l c | c c c c c | c }
\toprule
\textbf{Rubric Criteria} & \textbf{Strat.} & \textbf{k=5} & \textbf{k=10} & \textbf{k=20} & \textbf{k=25} & \textbf{k=30} & \textbf{Pooled (90)} \\
\midrule
\multicolumn{8}{l}{\textit{\textbf{Scenario A Objective Constraints}}} \\
\midrule
$Q_8$: Word count $\le$ 200 & N & 33.42 & 26.82 & 22.18 & 21.40 & 21.17 & \textbf{19.38} \\
$Q_2$: Exact keys check & A & 78.15 & 75.04 & 72.62 & 72.74 & 72.56 & \textbf{71.90} \\
$Q_3$: Active voice check & Q & 83.30 & 82.45 & 81.58 & 81.80 & 81.71 & \textbf{81.77} \\
$Q_7$: Valid JSON & A & 69.12 & 71.10 & 76.96 & 78.29 & 80.43 & \textbf{87.78} \\
$Q_1$: First sentence checks (Name/Period) & Q & 90.30 & 89.12 & 88.64 & 88.46 & 87.89 & \textbf{88.36} \\
$Q_6$: Project examples count $\ge$ 2 & Q & 94.02 & 94.55 & 95.38 & 94.76 & 94.95 & \textbf{95.87} \\
$Q_4$: Verbatim strength match & Q & 98.15 & 97.39 & 97.66 & 97.75 & 97.78 & \textbf{97.53} \\
$Q_5$: Growth area specific phrase & Q & 98.90 & 99.21 & 99.22 & 99.02 & 99.34 & \textbf{99.21} \\
\midrule
\multicolumn{8}{l}{\textit{\textbf{Scenario B Objective Constraints}}} \\
\midrule
$Q_8$: Vision statement word count & N & 39.71 & 35.71 & 32.55 & 31.23 & 30.76 & \textbf{29.84} \\
$Q_3$: Final sentence future tense & Q & 79.00 & 76.92 & 75.72 & 74.85 & 74.56 & \textbf{74.03} \\
$Q_1$: First-person pronoun check & A & 87.00 & 88.17 & 87.82 & 88.94 & 87.88 & \textbf{88.15} \\
$Q_5$: Contains $\ge$ 2 skills & Q & 92.17 & 92.06 & 91.77 & 92.95 & 91.77 & \textbf{92.34} \\
$Q_4$: Matches role AND job interest & Q & 95.38 & 95.31 & 94.21 & 93.99 & 94.38 & \textbf{94.08} \\
$Q_7$: Valid JSON, keys, and no markdown & A & 94.83 & 95.08 & 95.09 & 95.19 & 95.37 & \textbf{95.19} \\
$Q_6$: Exact match for career goals & Q & 96.92 & 96.94 & 96.85 & 96.67 & 96.24 & \textbf{96.40} \\
$Q_2$: Headline starts with active verb & Q & 97.67 & 97.36 & 97.27 & 97.13 & 97.54 & \textbf{97.15} \\

\bottomrule
\end{tabular}
\end{table*}

\subsubsection{Objective Reasoning Stability: The Determinism of Retrieval}
Table~\ref{tab:objective_reasoning_consolidated} presents the reasoning stability for all Objective constraints across both scenarios. The data reveals three distinct stability tiers driven by the specific \textbf{nature of the verification task}, whether it required continuous quantification, discrete retrieval, or structural validation.

 \textbf{Continuous Quantification (Content Failure):} As evidenced by the ``Word Count'' criteria $Q_8$ in both Scenario A ($19.38\%$) and Scenario B ($29.84\%$), LLMs consistently fail to reproduce \textbf{numeric} estimates for continuous data such as word counts. This confirms the ``Stability Trap'':  while the binary verdicts were nearly unanimous ($>99\%$), the reasoning process failed. Although the judges consistently employed a numeric strategy, the evidence space was unconstrained due to the opacity of tokenization. Lacking a deterministic mechanism to count words, the judges hallucinated inconsistent integer values (e.g., `185' vs `192') to retroactively justify their decisions.

\noindent \textbf{Discrete Content Retrieval (Stable Extraction):} Conversely, when the quantitative task was shifted to counting discrete entities (Scenario A $Q_6$: Project Examples and Scenario B $Q_5$: Skills), the stability increased above $90\%$. The nature of the task constrained the evidence space to extractable strings present in the input text. Consequently, the judges converged on a \textbf{quote strategy} with more consistent evidence across runs compared to continuous tasks.

\noindent \textbf{Structural Validation (Strategy Oscillation).} Comparing structural checks reveals how the specific mechanics of verification influences strategy stability. The atomic key check in Scenario A ($Q_2$) yielded moderate stability ($71.90\%$). We hypothesize that the simplicity of the task allowed judges to \textbf{oscillate strategy} between enumerating specific keys (e.g., ``Contains `name' '') and providing summary confirmations (e.g., ``All required keys are present''). In contrast, purely syntactic validations such as the ``Valid JSON'' checks in Scenario A ($Q_7$, $87.78\%$) and the complex structural check in Scenario B ($Q_7$, $95.19\%$) achieved higher stability. We conjecture that verifying syntax (as opposed to item enumeration) naturally constrains the judges to a `summary \textbf{assertion}' strategy, effectively converging the reasoning traces onto a single stable path of high-level abstraction.

\subsubsection{Subjective Reasoning Stability: The Spectrum of Extraction}
Table~\ref{tab:subjective_reasoning_consolidated} aggregates the subjective criteria. The results demonstrate that subjective instability is not random, but is driven by the ambiguity of the evidence target. Specifically, we observed that stability collapsed whenever the task implicitly required granular extraction from a fuzzy search space, whereas tasks that allowed for high-level abstraction remained moderately robust.

\begin{table*}[htbp]
\centering
\small
\renewcommand{\arraystretch}{1.2}
\setlength{\tabcolsep}{6pt}
\caption{\textbf{Subjective Reasoning Stability (Both Scenarios)}. Sorted by Pooled Stability. \textbf{Scenario} A=HR Review, B=Career Vision. \textbf{Strat.} denotes the rationale extraction strategy from justifications: (N)umeric, (Q)uote, (A)ssertion.}
\label{tab:subjective_reasoning_consolidated}
\begin{tabular}{l c | c c c c c | c }
\toprule
\textbf{Rubric Criteria} & \textbf{Strat.} & \textbf{k=5} & \textbf{k=10} & \textbf{k=20} & \textbf{k=25} & \textbf{k=30} & \textbf{Pooled (90)} \\
\midrule
\multicolumn{8}{l}{\textit{\textbf{Scenario A Subjective Constraints}}} \\
\midrule
$Q_7$: Encourages improvement & A & 39.42 & 36.74 & 35.73 & 36.49 & 36.90 & \textbf{40.08} \\
$Q_5$: Avoids absolute terms & A & 49.30 & 46.09 & 44.87 & 45.08 & 45.70 & \textbf{46.52} \\
$Q_2$: Clear and direct feedback & A & 57.80 & 57.30 & 56.52 & 57.56 & 57.55 & \textbf{58.90} \\
$Q_3$: Personality traits & A & 61.42 & 59.32 & 57.50 & 57.12 & 57.38 & \textbf{59.29} \\
$Q_6$: Frames areas as opportunities & A & 65.18 & 63.54 & 64.64 & 65.10 & 65.23 & \textbf{68.71} \\
$Q_1$: Constructive tone & A & 61.55 & 61.19 & 64.45 & 65.70 & 66.54 & \textbf{70.90} \\
$Q_4$: Avoids corporate jargon & A & 69.22 & 68.95 & 69.54 & 70.02 & 71.55 & \textbf{76.04} \\
\midrule
\multicolumn{8}{l}{\textit{\textbf{Scenario B Subjective Constraints}}} \\
\midrule
$Q_7$: Action-oriented vision & A & 41.83 & 35.54 & 34.66 & 34.70 & 35.28 & \textbf{35.24} \\
$Q_4$: Professional yet accessible & A & 42.50 & 39.33 & 36.72 & 36.88 & 37.06 & \textbf{36.07} \\
$Q_3$: Motivational tone & A & 49.21 & 46.10 & 45.10 & 44.23 & 45.29 & \textbf{46.77} \\
$Q_2$: Re-frame negatives & A & 53.08 & 50.88 & 50.01 & 51.28 & 51.01 & \textbf{53.04} \\
$Q_5$: Demonstrates empathy & A & 52.17 & 49.90 & 49.29 & 49.03 & 49.72 & \textbf{53.20} \\
$Q_1$: Cohesive narrative & A & 69.67 & 68.58 & 69.21 & 69.55 & 68.42 & \textbf{70.44} \\
$Q_6$: Punchy slogan headline & A & 83.92 & 82.94 & 82.98 & 82.77 & 81.90 & \textbf{83.26} \\
\bottomrule
\end{tabular}
\end{table*}

\noindent \textbf{Fuzzy Extraction (Stochastic Selection):} The lowest stability scores ($\approx 35\%-45\%$) occurred whenever the task implicitly required the judges to isolate specific words to prove a subjective trait. Whether searching for ``action-oriented'' verbs (Scenario B $Q_7$) or ``absolute terms'' (Scenario A $Q_5$), the lack of a defined search space caused judges to select different evidence across runs. Consequently, the judges failed to converge on a consistent evidence set, citing different examples across runs (e.g., ``transform'' vs ``lead'') to support the same verdict. This mirrors the failure seen in objective quantification: without a constrained target, the evidence selection becomes stochastic.

\noindent \textbf{Holistic Classification (Stable Abstraction):} Stability improved significantly ($>70\%$) for criteria where the task allowed for holistic abstraction. For tasks such as ``Avoids Corporate Jargon'' (Scenario A $Q_4$) or evaluating ``a Punchy Slogan'' (Scenario B $Q_6$) the reasoning traces consistently converged on broad summary assertions rather than attempting granular linguistic forensics. This confirms that subjective evaluation is reliable only when the justification strategy abstracts away from the noisy, token-level evidence space.

\section{Discussion}\label{sec:discussion}

Our findings reveal a critical paradox: while ``LLM-as-a-Judge'' frameworks are adopted for scalability, they exhibit a \textit{Stability Trap}, where high surface-level verdict agreement masks reasoning stability failures. We note that these findings are not issues that we have observed in deployed [Company] products or systems.

\paragraph{The Anthropomorphic Fallacy and the Stability Trap.}
We posit that the misuse of LLMs for objective tasks stems from an \textit{Anthropomorphic Fallacy}~\cite{reinecke2025double}: because models excel at complex semantic tasks (e.g., writing sonnets), operators assume they must be capable of trivial arithmetic (e.g., counting words). While prompting is easier than coding, it introduces stochastic failure modes invisible without reasoning analysis.
Most concerning is \textit{how} judges fail: they often hallucinate evidence to retroactively support a ``compliant'' verdict. This creates a worst-case scenario where a ``Trapped Judge'' (consistent Yes verdict, hallucinated math) bypasses human review entirely, allowing non-compliant artifacts to enter production. Unlike an unstable judge that acts as a ``canary in the coal mine'' by flipping verdicts, a trapped judge generates a corrupted audit trail that offers an illusion of safety.

\paragraph{A Hybrid Governance Architecture.}
To mitigate this, we propose bifurcating verification based on the cognitive nature of the task rather than just the tool's convenience.
\begin{enumerate*}
    \item \textbf{Tier 1: Deterministically Verifiable Constraints (Code-First).}  Any constraint that can be mathematically verified, including quantitative limits (e.g., word counts), structural validation (e.g., JSON schema), and exact substring matching, should be enforced via deterministic code. 

    \item \textbf{Tier 2: Semantic and Stylistic Evaluation (LLM-First).} LLM judges should be reserved exclusively for constraints beyond regular expressions, such as evaluating tone, reading comprehension, or complex grammatical structures (e.g., active voice). 

\end{enumerate*}

\paragraph{Implications for Design and Regulation.}
Product teams must adopt a strategy of \textit{Design for Auditability}, prioritizing structured formats (e.g., JSON schemas) over natural language instructions. This allows auditors to offload all objective verification to code. For example, rather than asking an LLM judge ``Did the output mention `Project X'?'', auditors should utilize exact string search or regex scripts. For regulated sectors (Finance, HR), this has legal weight. Emerging frameworks like the EU AI Act~\cite{EU_AI_Act_2024} mandate meaningful oversight for high-risk AI systems, requiring human-centered monitoring~\cite{shergadwala2022human}. If an automated judge records a compliant verdict based on hallucinated evidence, it generates a ``corrupted'' audit trail, misleading human reviewers. Consequently, the definition of ``audit stability'' must evolve from simple verdict consistency to \textit{semantic faithfulness}, ensuring the justification is factually grounded.

\paragraph{Limitations and Future Work.}

Our study primarily investigates the stability of the ``LLM-as-a-Judge'' framework, specifically the consistency of instruction adherence verdicts and the accompanying justifications, rather than its accuracy against a gold standard. While we acknowledge that high stability does not guarantee correctness (a model can be consistently wrong), we posit that stability is a fundamental prerequisite for automated governance.

We also recognize that the performance of any LLM is inextricably linked to the specific phrasing of its system prompt. While we adhered to current best practices for prompt engineering, we did not perform an exhaustive prompt variation study. Techniques such as few-shot prompting or chain-of-thought optimization are likely to influence the stability scores reported here. Furthermore, our experiments were conducted under relatively short context windows. Future work should evaluate the degradation of adherence stability in long-context scenarios (such as rubrics with exceeding 10 criteria), where ``lost-in-the-middle''~\cite{liu2024lost} phenomena might further exacerbate the instability we observed.

Methodologically, we treat verifiability as a binary distinction between Objective and Subjective tasks, yet we acknowledge this is often a spectrum. Our proposed HITL protocol~\ref{subsec:hitl} relies on the judgment of SMEs to correctly classify edge cases during the design phase, introducing a dependency on human expertise that may vary across organizations. Also, conditional instructions were out of the scope of this study based on reference-free audit. Future work can leverage our reasoning stability approach to further investigate LLM reasoning styles and audit requirements.

\section{Conclusion}\label{sec:conclusion}

This study challenges the assumption that ``instability'' is the primary barrier to automated AI governance. Instead, we identify a more insidious phenomenon: the \textit{Stability Trap}, where high surface-level agreement on adherence tasks masks deep reasoning failures. Our empirical analysis demonstrates that while LLM judges can achieve near-perfect verdict agreement, this stability often relies on hallucinated evidence, particularly in continuous quantification tasks (e.g., word counting), rendering it fragile for high-stakes compliance. Conversely, we found that modern models exhibit robust verdict \textit{and} reasoning stability on subjective stylistic criteria and discrete semantic extraction.

We conclude that the path to scalable governance lies in a Hybrid Architecture that delegates \textbf{all deterministically verifiable logic} to code while reserving LLM intelligence for \textbf{complex semantic evaluation}. For regulated domains like HR and Finance, this decoupling is not merely an optimization; it is a prerequisite for audit integrity. Future work should extend this analysis to Agentic AI workflows, evaluating whether the reasoning instability observed here propagates to autonomous tool-use and decision-making capabilities.

\section*{Ethical Considerations}
Given the sensitive nature of Human Resources and Talent Management, this study prioritized data privacy by utilizing a structurally faithful synthetic data generation approach. No proprietary customer data or Personally Identifiable Information (PII) regarding real workers was processed or exposed. This approach mitigates the risks of data leakage and ensures that our study does not inadvertently harm human subjects. Furthermore, the motivation of this work stems from mitigating potential downstream impact of the highlighted results in a production HR system which could in principle impact workers’ long‑term opportunities if such issues were left unaddressed in real‑world systems; this motivates our recommendation for Hybrid Automated Auditing Protocols to ensure worker protection. We also note that this research is intended to complement existing fairness, non‑discrimination, and compliance controls in enterprise HR systems by strengthening the reliability of automated evaluations; it does not assess or imply any disparate impact or bias in deployed systems.

\section*{Generative AI usage statement}
Gemini 3 Pro was utilized for rewriting original content written by the author to check for grammar, help with wordsmithing and reducing content to fit page limit. The authors reviewed and take full responsibility for the final content.


\section*{Acknowledgments}
We extend our gratitude to Sebastian Fuentes for his technical support in executing the data generation pipelines, including the synthesis of input variables and the orchestration of the AUT output generation across the four target architectures. We also thank Jason Vantomme, Veena Calambur, and several other Subject Matter Experts including the internal Legal and Compliance teams for their guidance on the HR use case definitions and regulatory constraints. Finally, we acknowledge the engineering teams who supported the infrastructure for the model deployments. 


\section*{Positionality Statement}
The authors approach this work from the perspective of Responsible AI practitioners in the enterprise software industry. While the software discussed is deployed globally, our analysis primarily centers on regulatory frameworks and workforce norms characteristic of the Global North (specifically North America and Europe).




\bibliographystyle{ACM-Reference-Format}
\bibliography{references}

\appendix

\section{LLM Prompt for Objective Rubric Generation}
\label{appendix:rubricGenObj}

The following prompt was used to extract Objective Non-Conditional Instructions ($I_{obj}$). The LLM is explicitly instructed to ignore subjective or conditional constraints.

\begin{figure*}[ht]
\centering
\begin{tcolorbox}[colback=gray!5!white,colframe=blue!50!black,title=Objective Rubric Generation Prompt]
\small
\begin{verbatim}
You are a meticulous Quality Assurance Engineer. Your primary task is to transform 
a user's instructions for an LLM into a simple, objective, yes/no checklist.

Your goal is to create questions that can be used to automatically evaluate an 
LLM's output. The user's original instructions may be subjective or contain 
conditional logic, but **your generated questions must always be objective and 
non-conditional.**

---
### Core Rules and Logic

**1. How to Handle Objective & Non-Conditional Instructions:**
  - If an instruction is objective and applies in all cases, transform it into a 
    direct yes/no question.
  - **Instruction -> Question Examples:**
    - Instruction: "The output must be 100 words long."
      -> Question: "Is the output exactly 100 words long?"
    - Instruction: "Format the entire response as a JSON object."
      -> Question: "Is the output a valid JSON object?"

**2. How to Handle Subjective or Conditional Instructions:**
  - If an instruction requires subjective opinion or is conditional, **do not 
    create a question for it.**
  - Instead, add the original instruction text to the "Ignored Instructions" list.
  - **Instruction -> Action Examples:**
    - Instruction: "Write in a friendly and polite tone."
      -> Action: Add to "Ignored Instructions" with reason "Subjective".
    - Instruction: "If the input mentions a city, include a fun fact."
      -> Action: Add to "Ignored Instructions" with reason "Conditional".

**3. Be Granular:**
  - If a single instruction applies to multiple items (such as "Ensure each paragraph 
    is under 50 words"), generate a separate question for each item.

**4. Group by Category:**
  - Organize your generated questions into: "Instruction Completeness", "Data 
    Field Utilization", "Format Adherence", and "Style Adherence".

---
### Output Format and Example
[Full JSON example structure omitted for brevity, see Appendix text]
\end{verbatim}
\end{tcolorbox}
\caption{The system prompt used to extract Objective Non-Conditional Instructions.}
\Description[System prompt for generating the Objective Rubric]{The figure displays the "Objective Rubric Generation Prompt" enclosed in a text box. The prompt instructs the AI model to adopt the persona of a "meticulous Quality Assurance Engineer." Its goal is to transform user instructions into a "simple, objective, yes/no checklist." 

The prompt outlines four "Core Rules and Logic":
1. **Handle Objective and Non-Conditional Instructions:** Convert these into direct yes/no questions (such as changing "The output must be 100 words long" to "Is the output exactly 100 words long?").
2. **Handle Subjective or Conditional Instructions:** Explicitly do not create questions for these. Instead, add them to an "Ignored Instructions" list with reasons such as "Subjective" (such as tone constraints) or "Conditional" (such as if/then logic).
3. **Be Granular:** Generate separate questions for repeated items.
4. **Group by Category:** Organize questions into specific categories like "Instruction Completeness" and "Format Adherence."}
\label{fig:prompt_obj}
\end{figure*}

\newpage

\section{LLM Prompt for Subjective Rubric Generation}
\label{appendix:rubricGenSubj}

The following prompt was used to extract Subjective Non-Conditional Instructions ($I_{subj}$). Note the ``CRITICAL SCOPE'' instruction that inverts the logic of the objective prompt, targeting tone and style while ignoring hard constraints.

\begin{figure*}[htbp]
\centering
\begin{tcolorbox}[colback=gray!5!white,colframe=red!50!black,title=Subjective Rubric Generation Prompt]
\small
\begin{verbatim}
You are a meticulous Quality Assurance Engineer. Your primary task is to transform 
a user's instructions for an LLM into a simple, objective, yes/no checklist.

**CRITICAL SCOPE:** You must focus ONLY on instructions that are **Subjective** AND **Non-Conditional**.

---
### Core Rules and Logic

**1. How to Handle Subjective & Non-Conditional Instructions (TARGET):**
  - If an instruction requires human judgment, opinion, or refers to tone, style, 
    creativity, or "soft skills," AND it applies in all cases (no "if/when"), 
    transform it into a yes/no question.
  - **Instruction -> Question Examples:**
    - Instruction: "Write in a friendly and polite tone."
      -> Question: "Is the tone of the response friendly and polite?"
    - Instruction: "The story should be engaging and dramatic."
      -> Question: "Is the story engaging and dramatic?"

**2. How to Handle Objective or Conditional Instructions (IGNORE):**
  - **Objective:** If an instruction is a hard fact that can be measured with 
    code (exact word counts, specific data fields, JSON format), **do not create 
    a question.**
  - **Conditional:** If an instruction depends on logic, **do not create a 
    question.**
  - **Instruction -> Action Examples:**
    - Instruction: "The output must be exactly 100 words."
      -> Action: Add to "Ignored Instructions" with reason "Objective".
    - Instruction: "If the user is angry, be empathetic."
      -> Action: Add to "Ignored Instructions" with reason "Conditional".

**3. Group by Category:**
  - Organize your generated questions into: "Instruction Completeness", "Data 
    Field Utilization", "Format Adherence", and "Style Adherence".

---
### Output Format and Example
[Full JSON example structure omitted for brevity, see Appendix text]
\end{verbatim}
\end{tcolorbox}
\caption{The system prompt used to extract Subjective Non-Conditional Instructions.}
\Description[System prompt for generating the Subjective Rubric]{The figure displays the "Subjective Rubric Generation Prompt" enclosed in a text box. Similar to the objective prompt, it instructs the AI to act as a Quality Assurance Engineer but defines a "CRITICAL SCOPE" to focus ONLY on Subjective and Non-Conditional instructions.

The prompt outlines "Core Rules and Logic":
1. **Handle Subjective & Non-Conditional Instructions (TARGET):** Transform instructions requiring human judgment, opinion, tone, or style into yes/no questions (such as converting "Write in a friendly tone" to "Is the tone friendly?").
2. **Handle Objective or Conditional Instructions (IGNORE):** Explicitly ignore hard facts (Objective) and logic-gated rules (Conditional). Examples show adding instructions like "Output must be 100 words" to an "Ignored Instructions" list with the reason "Objective".
3. **Group by Category:** Organize the output into standard categories like "Style Adherence" and "Instruction Completeness."}
\label{fig:prompt_subj}
\end{figure*}

\section{LLM Judge Scoring Prompt}
\label{appendix:judgePrompt}

The following system prompt was used for the Judge LLM in Stage 3 of the framework. It enforces a strict, reference-free evaluation using only the provided rubric categories. The instruction to provide a ``brief justification'' is the mechanism used to induce reasoning described in Section~\ref{subsec:hitl}.

\begin{figure*}[htbp]
\centering
\begin{tcolorbox}[colback=gray!5!white,colframe=black!75!white,title=Judge Scoring Prompt]
\small
\begin{verbatim}
You are a strict and fair evaluator. Your task is to evaluate an AI-generated 
output based *only* on the provided categorized rubric.
For each question, you must provide a "Yes" or "No" answer and a brief 
justification.

Original Instructions start:
---
{filled_prompt}
---
Original Instructions End

AI-Generated Output to Evaluate Start:
---
{model_output}
---
AI-Generated Output to Evaluate End

Evaluation Rubric Start:
---
{categorized_rubric_json}
---
Evaluation Rubric End

Your output must be a JSON object. The main keys must be the exact rubric 
categories provided above.
Each category must contain a list of objects, where each object has:
- 'question': (Must match the rubric exactly)
- 'justification': (A brief explanation)
- 'answer': ("Yes" or "No")

Do not create any questions yourself.
Do not skip any questions from the rubric.
\end{verbatim}
\end{tcolorbox}
\caption{The system prompt used by the Judge LLM to evaluate adherence. It requires a structured JSON output with mandatory justifications for every verdict.}
\Description[System prompt for the Judge LLM]{The figure displays the "Judge Scoring Prompt" enclosed in a text box. It directs the AI to act as a "strict and fair evaluator" whose task is to judge an AI-generated output based *only* on a provided rubric. 

The prompt structure includes placeholders for three inputs:
1. `{filled_prompt}` (The Original Instructions)
2. `{model_output}` (The AI Output to Evaluate)
3. `{categorized_rubric_json}` (The Evaluation Rubric)

It mandates a specific JSON output format where each item must include the exact question from the rubric, a brief text justification, and a "Yes" or "No" answer. The prompt explicitly restricts the model from creating new questions or skipping any existing ones.}
\label{fig:prompt_judge}
\end{figure*}

\section{AUT System Prompt: Scenario A}
\label{appendix:appPromptA}

The following system prompt was used to instruct the Generative Agent (LLM) in Scenario A: HR Performance Review. It contains a mix of Objective constraints (such as JSON format, exact string inclusion) and Subjective constraints (such as constructive tone, avoiding jargon) that form the basis of our stability analysis.

\begin{figure*}[htbp]
\centering
\begin{tcolorbox}[colback=gray!5!white,colframe=black!75!white,title=Application Prompt: Scenario A (HR Business Partner)]
\small
\begin{verbatim}
You are an experienced HR Business Partner assisting a manager in writing a 
performance review feedback. Your goal is to synthesize the provided data into 
a clear and balanced summary.

**Input Data:**
* Employee name: `{employee_name}`
* Review period: `{review_period}`
* Key strengths: `{key_strengths}`
* Areas for growth: `{areas_for_growth}`
* Project examples: `{project_examples}`
* Promotion Readiness: `{is_promotion_ready}`

**Action:** Generate a performance review summary for `{employee_name}`.

**Guidelines:**
You must adhere to the following guidelines in your response. The final output 
must be a valid JSON object with two keys: "strengths_summary" and 
"growth_areas_summary". You must ensure that the text value of the 
"strengths_summary" key begins immediately with the employee's name 
(`{employee_name}`) and the `{review_period}`.

Throughout the summary, maintain a constructive and professional tone. **You must 
use active voice to ensure the feedback is clear and direct. Strictly avoid 
corporate jargon (e.g., "synergy", "drill down") or absolute terms like "always" 
and "never" which can sound biased.** You must generate the "strengths_summary" 
containing the **exact text** of at least one item from `{key_strengths}`. For 
the "growth_areas_summary", you must **begin** the sentence with the strict 
phrase **'To address [Insert Exact Item from `{areas_for_growth}`],'** and then 
complete the sentence by crafting an encouraging recommendation for how the 
employee can improve that specific skill. Your writing should focus on observable 
behaviors and measurable results rather than personality traits. Be sure to 
incorporate at least two specific examples from the `{project_examples}` list 
**verbatim** into the text, and keep the total length of the combined summaries 
to no more than 200 words.

If `{is_promotion_ready}` is "Yes", you must add a sentence to the 
"strengths_summary" discussing the employee's readiness for future challenges 
or a greater scope of responsibility. If the `{areas_for_growth}` list is empty, 
the value for the "growth_areas_summary" key must be the exact string: "No 
significant areas for growth were identified in this period." To conclude, 
craft a motivational and encouraging closing sentence for the 
"strengths_summary" to inspire continued high performance.
\end{verbatim}
\end{tcolorbox}
\caption{The system prompt used to generate the HR Performance Review outputs for Scenario A.}
\Description[Application Prompt for Scenario A (HR Review)]{The figure displays the "Application Prompt: Scenario A (HR Business Partner)." It assigns the AI the persona of an experienced HR Business Partner tasked with synthesizing employee data (Name, Review Period, Strengths, Growth Areas, Project Examples, Promotion Readiness) into a performance review.

The prompt enforces strict guidelines mixing format, style, and content:
1. **Format:** Output must be a valid JSON object with keys `strengths_summary` and `growth_areas_summary`.
2. **Style:** Requires active voice, constructive tone, and explicitly bans corporate jargon (such as "synergy") and absolute terms (such as "always").
3. **Constraints:** Includes specific start strings (must begin with employee name), verbatim inclusion of input data (specific project examples), and a maximum word count of 200 words.
4. **Logic:** Contains conditional instructions for handling `is_promotion_ready` (adding specific sentences) and empty growth lists (using a fallback string).}
\label{fig:prompt_scenario_a}
\end{figure*}

\section{AUT System Prompt: Scenario B}
\label{appendix:appPromptB}

The following system prompt was used to instruct the Generative Agent (LLM) in Scenario B: Career Vision Statement. This scenario introduces higher creative complexity (such as "motivational tone," "cohesive narrative arc") while retaining strict structural constraints (such as JSON schema, word counts), providing a distinct test case for scoring stability compared to Scenario A.

\begin{figure*}[htbp]
\centering
\begin{tcolorbox}[colback=gray!5!white,colframe=black!75!white,title=Application Prompt: Scenario B (Career Vision Statement)]
\scriptsize
\begin{lstlisting}[breaklines=true, basicstyle=\ttfamily]
You are an expert Career Coach and Talent Development Specialist. Your goal is to synthesize a worker's professional profile into a compelling "Career Vision Statement." This statement will serve as a north star for the worker's development and a conversation starter for their mentors.

**Input Data:**
[Worker Name: `{worker_name}`, Current Role: `{current_role}`, Work History: `{work_history}`, Career Goals: `{career_goals}`, Top Skills: `{top_skills}`, Skill Interests: `{skill_interests}`, Development Items: `{development_items}`, Job Interests: `{job_interests}`, Recent Feedback: `{feedback}`]

**Action:** Generate a personalized Career Vision Statement for `{worker_name}`.

**Guidelines:**
You must adhere to the following instructions to generate the response.

1. **Strict JSON Formatting:** You must format the final output as a valid JSON object containing exactly two keys: "headline_summary" and "detailed_vision_statement". Do not include markdown formatting (like ```json) outside the object. To ensure the JSON is parseable, you must strictly follow these syntax rules inside the values:
   * **No physical line breaks:** You must escape all paragraphs using the characters "\\n" (double backslash n). Do not use actual newlines.
   * **No double quotes inside text:** You must strictly use single quotes (') for any internal quoting, emphasis, or titles. NEVER use double quotes (") inside the text values, as this breaks the JSON structure.

2. In both the "headline_summary" and "detailed_vision_statement", you must write the content using a first-person point of view (using "I", "my" or "me") to ensure the worker owns the statement.
3. You must strictly limit the "detailed_vision_statement" value to a maximum of 500 words.
4. In the "detailed_vision_statement", you must include the exact text of the `{current_role}` and at least one specific item from `{job_interests}` verbatim.
5. In the "detailed_vision_statement", you must incorporate at least two specific skills verbatim from either the `{skill_interests}` list or the `{top_skills}` list.
6. You must begin the "headline_summary" with an active verb describing a future state (e.g., "Building," "Leading," "Transforming").
7. You must cite one specific goal from `{career_goals}` verbatim within the "detailed_vision_statement".
8. In both the "headline_summary" and "detailed_vision_statement", you must ensure the overall tone is highly motivational and aspirational to inspire the worker during difficult times.
9. In the "detailed_vision_statement", you must synthesize the `{work_history}` and `{career_goals}` to create a cohesive narrative arc that bridges where they have been with where they are going.
10. In the "detailed_vision_statement", you must frame the `{development_items}` and `{feedback}` positively, focusing on the potential for mastery rather than the deficit of the skill.
11. In both the "headline_summary" and "detailed_vision_statement", you must use professional yet accessible language that is suitable for sharing with senior leadership or mentors.
12. In the "detailed_vision_statement", you must demonstrate empathy by reflecting the worker's passion as implied by their `{job_interests}` and `{skill_interests}`.
13. You must ensure the "headline_summary" acts as a punchy, memorable personal brand slogan.
14. In the "detailed_vision_statement", you must craft the vision to feel "action-oriented," creating a sense of momentum and immediate relevance to their career path.
15. The final sentence of the 'detailed_vision_statement' must be written in the future tense.
**Output Examples:**
POSITIVE EXAMPLE (Do this - Valid Syntax):
[Full JSON example structure omitted for brevity]
WRONG EXAMPLE (Do NOT do this - Invalid Syntax):
[Full JSON example structure omitted for brevity]
\end{lstlisting}
\end{tcolorbox}
\caption{The system prompt used to generate the Career Vision Statement outputs for Scenario B. Note the explicit constraints regarding JSON syntax, escaping, and quoting.}
\label{fig:prompt_scenario_b}
\Description[Application Prompt for Scenario B (Career Vision Statement)]{The figure displays the "Application Prompt: Scenario B (Career Vision Statement)." It assigns the AI the persona of an expert Career Coach synthesizing extensive worker data (Name, Role, History, Goals, Skills, Feedback, etc.) into a Career Vision Statement.

The prompt enforces highly specific and complex constraints:
1. **Strict JSON Syntax:** The output must be a valid JSON object with keys `headline_summary` and `detailed_vision_statement`. It explicitly mandates using escaped `\n` for line breaks and single quotes (') for internal text, banning physical newlines and double quotes to ensure parseability.
2. **Content Constraints:** Requires first-person POV, a 500-word maximum, verbatim inclusion of specific roles, skills, and goals, an active-verb headline start, and a future-tense final sentence.
3. **Tone & Synthesis:** Demands a motivational, action-oriented narrative that synthesizes history with future goals and frames negative feedback positively.
The prompt includes explicit positive and negative examples to illustrate the required JSON syntax.}
\end{figure*}

\section{Evaluation Rubrics}\label{appendix:evalrubrics}

This appendix details the complete sets of Objective and Subjective criteria used to evaluate instruction adherence for both experimental scenarios. These rubrics were generated using the hybrid HITL protocol described in Section~\ref{subsec:hitl} and classified according to the SID Taxonomy.

\begin{table*}[htbp]
\caption{Scenario A (HR Performance Review) - Objective Evaluation Criteria ($I_{obj}$)}
\label{tab:scenario_a_objective}
\centering
\begin{tabular}{@{}p{0.25\linewidth}p{0.7\linewidth}@{}}
\toprule
\textbf{Dimension} & \textbf{Criteria (Binary Verification)} \\ \midrule
\textbf{Instruction Completeness} 
& 1. Does the first sentence of \texttt{strengths\_summary} contain both \texttt{\{employee\_name\}} and \texttt{\{review\_period\}}? \\
& 2. Does the output contain exactly two keys: \texttt{strengths\_summary} and \texttt{growth\_areas\_summary}? \\
& 3. Does the text value of \texttt{strengths\_summary} use active voice? \\ \midrule
\textbf{Data Field Utilization} 
& 4. Does \texttt{strengths\_summary} contain a verbatim substring match from \texttt{\{key\_strengths\}}? \\
& 5. Does \texttt{growth\_areas\_summary} begin with ``To address '' followed by an exact item from \texttt{\{areas\_for\_growth\}}? \\
& 6. Does the output contain at least two exact string matches from \texttt{\{project\_examples\}}? \\ \midrule
\textbf{Format Adherence} 
& 7. Is the entire output a valid JSON object? \\
& 8. Is the combined word count of both values less than or equal to 200 words? \\ \bottomrule
\end{tabular}
\end{table*}

\begin{table*}[htbp]
\caption{Scenario A (HR Performance Review) - Subjective Evaluation Criteria ($I_{subj}$)}
\label{tab:scenario_a_subjective}
\centering
\begin{tabular}{@{}p{0.25\linewidth}p{0.7\linewidth}@{}}
\toprule
\textbf{Dimension} & \textbf{Criteria (Qualitative Assessment)} \\ \midrule
\textbf{Style Adherence} 
& 1. Does the output have a constructive tone? \\
& 2. Is the feedback clear and direct? \\
& 3. Does the strengths summary account for personality traits? \\
& 4. Does the output avoid corporate jargon? \\
& 5. Does the output avoid absolute terms (e.g., ``always'', ``never'')? \\
& 6. Does the output frame development areas as opportunities? \\
& 7. Does the style encourage future improvement? \\ \bottomrule
\end{tabular}
\end{table*}

\begin{table*}[htbp]
\caption{Scenario B (Career Vision Statement) - Objective Evaluation Criteria ($I_{obj}$)}
\label{tab:scenario_b_objective}
\centering
\begin{tabular}{@{}p{0.25\linewidth}p{0.7\linewidth}@{}}
\toprule
\textbf{Dimension} & \textbf{Criteria (Binary Verification)} \\ \midrule
\textbf{Instruction Completeness} 
& 1. Do both the \texttt{headline\_summary} and \texttt{detailed\_vision\_statement} text values contain at least one first-person pronoun? \\
& 2. Does the text value of the \texttt{headline\_summary} key contain an active verb that describes a future state? \\ 
& 3. Is the final sentence of the value of the key \texttt{detailed\_vision\_statement} written in the future tense? \\ \midrule
\textbf{Data Field Utilization} 
& 4. Does the \texttt{detailed\_vision\_statement} contain the exact string match for \texttt{\{current\_role\}} AND at least one exact match from the \texttt{\{job\_interests\}} list? \\
& 5. Does the \texttt{detailed\_vision\_statement} contain exact string matches for at least two items found in either the \texttt{\{skill\_interests\}} list or the \texttt{\{top\_skills\}} list? \\
& 6. Does the \texttt{detailed\_vision\_statement} contain an exact string match for the text provided in \texttt{\{career\_goals\}}? \\ \midrule
\textbf{Format Adherence} 
& 7. Is the final output strictly a valid JSON object containing exactly the keys \texttt{headline\_summary} and \texttt{detailed\_vision\_statement} without any markdown formatting? \\
& 8. Is the word count of the \texttt{detailed\_vision\_statement} value less than or equal to 500 words? \\ \bottomrule
\end{tabular}
\end{table*}

\begin{table*}[htbp]
\caption{Scenario B (Career Vision Statement) - Subjective Evaluation Criteria ($I_{subj}$)}
\label{tab:scenario_b_subjective}
\centering
\begin{tabular}{@{}p{0.25\linewidth}p{0.7\linewidth}@{}}
\toprule
\textbf{Dimension} & \textbf{Criteria (Qualitative Assessment)} \\ \midrule
\textbf{Instruction Completeness} 
& 1. Does the \texttt{detailed\_vision\_statement} successfully synthesize the \texttt{\{work\_history\}} and \texttt{\{career\_goals\}} into a cohesive narrative arc, rather than just listing them sequentially? \\
& 2. Are the \texttt{\{development\_items\}} and \texttt{\{feedback\}} re-framed strictly as positive opportunities for mastery, effectively avoiding any negative or deficit-based language? \\ \midrule
\textbf{Style Adherence} 
& 3. Is the overall tone of the response highly motivational and aspirational, suitable for inspiring the worker? \\
& 4. Is the language used in the response professional yet accessible enough to be shared with senior leadership? \\
& 5. Does the response demonstrate empathy by clearly reflecting the passion implied in the worker's interest fields? \\
& 6. Does the \texttt{headline\_summary} function effectively as a punchy and memorable personal brand slogan? \\
& 7. Does the vision statement feel ``action-oriented,'' conveying a clear sense of momentum and immediate relevance? \\ \bottomrule
\end{tabular}
\end{table*}

\section{Detailed Verdict Reliability Results}\label{appendix:verdict_detailed_results}

\subsection{Scenario A: Objective Verdict Stability}
Table~\ref{tab:scenario_a_objective_verdict} details the item-level stability for the HR Performance Review objective criteria. \textbf{Each cell represents the macro-average of stability metrics calculated independently for the 16 distinct Judge-Generator pairs.} The rows denote the number of repeated runs ($k$) used to compute these metrics, ranging from subsets ($k=5$) to the full experimental dataset (\textbf{Pooled (90)}).

The data confirms the high reliability of LLM judges for rigid syntactic tasks, with $100\%$ agreement observed across all 16 pairs for exact key verification ($Q_2$), active voice checks ($Q_3$), verbatim string match ($Q_4$), Project examples ($Q_6$), and JSON formatting ($Q_7$).

\begin{table*}[htbp]
\centering
\small 
\setlength{\tabcolsep}{2.5pt} 
\renewcommand{\arraystretch}{1.3} 

\newcommand{\res}[3]{\makecell{\textbf{#1} \\ (#2, #3)}}
\caption{\textbf{Verdict Stability Analysis (Scenario A: Objective)}. 
Values formatted as: \textbf{Percentage Agreement, (Gwet's AC1, Fleiss' Kappa)}.
Columns $Q_1$--$Q_8$ represent atomic question stability, while \textit{Rubric $\sigma$} represents the standard deviation of the aggregate adherence score.
\textbf{Key (mapped to Table~\ref{tab:scenario_a_objective}):} 
$Q_1$: First sentence checks (Name/Period); 
$Q_2$: Exact keys check; 
$Q_3$: Active voice check; 
$Q_4$: Verbatim strength match; 
$Q_5$: Growth area specific start phrase; 
$Q_6$: Verbatim Projects Match $\ge$ 2 ; 
$Q_7$: Valid JSON; 
$Q_8$: Word count $\le$ 200.}
\label{tab:scenario_a_objective_verdict}
\resizebox{\textwidth}{!}{%
\begin{tabular}{l c c c c c c c c | c}
\toprule
\textbf{Runs ($k$)} & \textbf{$Q_1$} & \textbf{$Q_2$} & \textbf{$Q_3$} & \textbf{$Q_4$} & \textbf{$Q_5$} & \textbf{$Q_6$} & \textbf{$Q_7$} & \textbf{$Q_8$} & \textbf{Rubric $\sigma$} \\
\midrule

$k=5$ & 
\res{99.72}{0.993}{0.987} & \res{100}{1.000}{1.000} & \res{100}{1.000}{1.000} & \res{100}{1.000}{1.000} & 
\res{99.85}{0.997}{0.990} & \res{100}{1.000}{1.000} & \res{100}{1.000}{1.000} & \res{99.75}{0.995}{0.922} & 
$0.1508$ \\ \addlinespace

$k=10$ & 
\res{99.76}{0.994}{0.989} & \res{100}{1.000}{1.000} & \res{100}{1.000}{1.000} & \res{100}{1.000}{1.000} & 
\res{99.75}{0.996}{0.986} & \res{100}{1.000}{1.000} & \res{100}{1.000}{1.000} & \res{99.65}{0.993}{0.892} & 
$0.1987$ \\ \addlinespace

$k=20$ & 
\res{99.72}{0.994}{0.988} & \res{100}{1.000}{1.000} & \res{100}{1.000}{1.000} & \res{100}{1.000}{1.000} & 
\res{99.79}{0.997}{0.989} & \res{100}{1.000}{1.000} & \res{100}{1.000}{1.000} & \res{99.54}{0.991}{0.808} & 
$0.2613$ \\ \addlinespace

$k=25$ & 
\res{99.72}{0.993}{0.988} & \res{100}{1.000}{1.000} & \res{100}{1.000}{1.000} & \res{100}{1.000}{1.000} & 
\res{99.78}{0.996}{0.987} & \res{100}{1.000}{1.000} & \res{100}{1.000}{1.000} & \res{99.53}{0.991}{0.862} & 
$0.2708$ \\ \addlinespace

$k=30$ & 
\res{99.66}{0.992}{0.985} & \res{100}{1.000}{1.000} & \res{100}{1.000}{1.000} & \res{100}{1.000}{1.000} & 
\res{99.81}{0.997}{0.988} & \res{100}{1.000}{1.000} & \res{100}{1.000}{1.000} & \res{99.56}{0.992}{0.873} & 
$0.3060$ \\ \addlinespace

\midrule
\textbf{Pooled (90)} & 
\res{99.70}{0.993}{0.987} & \res{100}{1.000}{1.000} & \res{100}{1.000}{1.000} & \res{100}{1.000}{1.000} & 
\res{99.77}{0.996}{0.988} & \res{100}{1.000}{1.000} & \res{100}{1.000}{1.000} & \res{99.57}{0.992}{0.810} & 
$0.3158$ \\

\bottomrule
\end{tabular}}
\end{table*}

\subsection{Scenario A: Subjective Verdict Stability}
Table~\ref{tab:scenario_a_subjective_verdict} presents the stability metrics for the HR Performance Review subjective criteria. \textbf{Similar to the objective analysis, these values represent the macro-average of stability metrics across the 16 distinct Judge-Generator pairs.}

The judges exhibited a distinct ``ceiling effect'' across nearly all dimensions, achieving absolute stability ($100\%$ verdict agreement) for 6 out of 7 criteria, including Constructive Tone ($Q_1$), Jargon Avoidance ($Q_4$), and Absolute Terms ($Q_5$). The only deviation observed was for Future Improvement ($Q_7$), though agreement remained exceptional ($>99.8\%$) across the pooled dataset. Consequently, the aggregate score volatility was negligible ($\sigma \approx 0.03$), indicating that for this specific task, the judges' subjective binary ratings were effectively static despite the potential for qualitative ambiguity.

\begin{table*}[htbp]
\centering
\small 
\setlength{\tabcolsep}{3.5pt} 
\renewcommand{\arraystretch}{1.3} 

\newcommand{\res}[3]{\makecell{\textbf{#1} \\ (#2, #3)}}

\caption{\textbf{Verdict Stability Analysis (Scenario A: Subjective)}. 
Values formatted as: \textbf{Percentage Agreement, (Gwet's AC1, Fleiss' Kappa)}. Note: Fleiss' Kappa is undefined (NaN) where agreement is 100\%.
Columns $Q_1$--$Q_7$ represent atomic question stability, while \textit{Rubric $\sigma$} represents the standard deviation of the aggregate adherence score.
\textbf{Key (mapped to Table~\ref{tab:scenario_a_subjective}):} 
$Q_1$: Constructive tone; 
$Q_2$: Clear and direct feedback; 
$Q_3$: Personality traits accounted for; 
$Q_4$: Avoids corporate jargon; 
$Q_5$: Avoids absolute terms; 
$Q_6$: Frames areas as opportunities; 
$Q_7$: Encourages future improvement.}
\label{tab:scenario_a_subjective_verdict}
\resizebox{\textwidth}{!}{%
\begin{tabular}{l c c c c c c c | c}
\toprule
\textbf{Runs ($k$)} & \textbf{$Q_1$} & \textbf{$Q_2$} & \textbf{$Q_3$} & \textbf{$Q_4$} & \textbf{$Q_5$} & \textbf{$Q_6$} & \textbf{$Q_7$} & \textbf{Rubric $\sigma$} \\
\midrule

$k=5$ & 
\res{100}{1.000}{1.000} & \res{100}{1.000}{1.000} & \res{100}{1.000}{1.000} & \res{100}{1.000}{1.000} & 
\res{100}{1.000}{1.000} & \res{100}{1.000}{1.000} & \res{99.85}{0.998}{0.796} & 
0.0355 \\ \addlinespace

$k=10$ & 
\res{100}{1.000}{1.000} & \res{100}{1.000}{1.000} & \res{100}{1.000}{1.000} & \res{100}{1.000}{1.000} & 
\res{100}{1.000}{1.000} & \res{100}{1.000}{1.000} & \res{99.89}{0.998}{0.853} & 
0.0245 \\ \addlinespace

$k=20$ & 
\res{100}{1.000}{1.000} & \res{100}{1.000}{1.000} & \res{100}{1.000}{1.000} & \res{100}{1.000}{1.000} & 
\res{100}{1.000}{1.000} & \res{100}{1.000}{1.000} & \res{99.84}{0.998}{0.821} & 
0.0336 \\ \addlinespace

$k=25$ & 
\res{100}{1.000}{1.000} & \res{100}{1.000}{1.000} & \res{100}{1.000}{1.000} & \res{100}{1.000}{1.000} & 
\res{100}{1.000}{1.000} & \res{100}{1.000}{1.000} & \res{99.89}{0.998}{0.796} & 
0.0286 \\ \addlinespace

$k=30$ & 
\res{100}{1.000}{1.000} & \res{100}{1.000}{1.000} & \res{100}{1.000}{1.000} & \res{100}{1.000}{1.000} & 
\res{100}{1.000}{1.000} & \res{100}{1.000}{1.000} & \res{99.91}{0.999}{0.786} & 
0.027 \\

\midrule
\textbf{Pooled (90)} & 
\res{100}{1.000}{1.000} & \res{100}{1.000}{1.000} & \res{100}{1.000}{1.000} & \res{100}{1.000}{1.000} & 
\res{100}{1.000}{1.000} & \res{100}{1.000}{1.000} & \res{99.88}{0.998}{0.806} & 
0.0304 \\

\bottomrule
\end{tabular}}
\end{table*}

\subsection{Scenario B: Objective Verdict Stability}
Table~\ref{tab:scenario_b_objective_verdict} details the results for the Career Vision Statement objective criteria. While the pooled agreement remained high ($>97\%$), this scenario exhibited the highest aggregate score volatility ($\sigma = 1.4176$) in the study. Notably, the Headline Active Verb Check ($Q_2$) demonstrates the ``Kappa Paradox'': despite a high percentage agreement of $99.80\%$, the Fleiss' Kappa score dropped to $0.523$ due to the skewed distribution of the binary labels, whereas Gwet's AC1 ($0.996$) correctly reflected the observed high consensus. Additionally, we observed perfect stability ($100\%$ agreement) for the Skills Match check ($Q_5$) and near-perfect stability for Word Count ($Q_8$), though the latter exhibited significant reasoning instability in the semantic analysis.

\begin{table*}[htbp]
\centering
\small 
\setlength{\tabcolsep}{3.5pt} 
\renewcommand{\arraystretch}{1.3} 

\newcommand{\res}[3]{\makecell{\textbf{#1} \\ (#2, #3)}}

\caption{\textbf{Verdict Stability Analysis (Scenario B: Objective)}. 
Values formatted as: \textbf{Percentage Agreement, (Gwet's AC1, Fleiss' Kappa)}.
Columns $Q_1$--$Q_8$ represent atomic question stability, while \textit{Rubric $\sigma$} represents the standard deviation of the aggregate adherence score.
\textbf{Key (mapped to Table~\ref{tab:scenario_b_objective}):} 
$Q_1$: First-person pronoun check; 
$Q_2$: Headline starts with active verb (future state); 
$Q_3$: Final sentence future tense; 
$Q_4$: Matches role AND job interest; 
$Q_5$: Matches $\ge$ 2 skills; 
$Q_6$: Exact match for career goals; 
$Q_7$: Valid JSON structure; 
$Q_8$: Word count $\le$ 500.}
\label{tab:scenario_b_objective_verdict}
\resizebox{\textwidth}{!}{%
\begin{tabular}{l c c c c c c c c | c}
\toprule
\textbf{Runs ($k$)} & \textbf{$Q_1$} & \textbf{$Q_2$} & \textbf{$Q_3$} & \textbf{$Q_4$} & \textbf{$Q_5$} & \textbf{$Q_6$} & \textbf{$Q_7$} & \textbf{$Q_8$} & \textbf{Rubric $\sigma$} \\
\midrule

$k=5$ & 
\res{99.79}{0.996}{0.810} & \res{99.92}{0.998}{0.874} & \res{98.79}{0.978}{0.772} & \res{99.08}{0.981}{0.898} & 
\res{100}{1.000}{1.000} & \res{97.83}{0.938}{0.786} & \res{99.67}{0.993}{0.765} & \res{100}{1.000}{1.000} & 
1.0382 \\ \addlinespace

$k=10$ & 
\res{99.69}{0.995}{0.835} & \res{99.73}{0.995}{0.712} & \res{98.58}{0.976}{0.765} & \res{99.10}{0.983}{0.807} & 
\res{100}{1.000}{1.000} & \res{97.58}{0.939}{0.782} & \res{99.88}{0.998}{0.750} & \res{100}{1.000}{1.000} & 
1.1506 \\ \addlinespace

$k=20$ & 
\res{99.57}{0.993}{0.841} & \res{99.78}{0.996}{0.639} & \res{98.51}{0.976}{0.768} & \res{99.01}{0.983}{0.805} & 
\res{100}{1.000}{1.000} & \res{97.80}{0.946}{0.731} & \res{99.78}{0.996}{0.726} & \res{99.99}{1.000}{0.937} & 
1.2399 \\ \addlinespace

$k=25$ & 
\res{99.59}{0.993}{0.839} & \res{99.82}{0.997}{0.583} & \res{98.62}{0.977}{0.768} & \res{99.02}{0.983}{0.812} & 
\res{100}{1.000}{1.000} & \res{97.45}{0.939}{0.728} & \res{99.77}{0.996}{0.789} & \res{99.99}{1.000}{0.937} & 
1.3100 \\ \addlinespace

$k=30$ & 
\res{99.64}{0.994}{0.829} & \res{99.81}{0.996}{0.576} & \res{98.65}{0.978}{0.759} & \res{98.99}{0.982}{0.798} & 
\res{100}{1.000}{1.000} & \res{97.74}{0.943}{0.717} & \res{99.84}{0.997}{0.796} & \res{99.99}{1.000}{0.937} & 
1.3387 \\

\midrule
\textbf{Pooled (90)} & 
\res{99.62}{0.994}{0.837} & \res{99.80}{0.996}{0.523} & \res{98.55}{0.977}{0.766} & \res{99.01}{0.983}{0.806} & 
\res{100}{1.000}{1.000} & \res{97.61}{0.942}{0.727} & \res{99.80}{0.996}{0.738} & \res{99.99}{1.000}{0.938} & 
1.4176 \\

\bottomrule
\end{tabular}
}
\end{table*}

\subsection{Scenario B: Subjective Verdict Stability}
Table~\ref{tab:scenario_b_subjective_verdict} presents the metrics for the Career Vision Statement subjective criteria. Consistent with Scenario A, the judges achieved near-perfect stability, with $100\%$ agreement observed for 6 out of 7 questions, including Cohesive Narrative ($Q_1$), Motivational Tone ($Q_3$), and Empathy ($Q_5$). The only minor deviation occurred in the Slogan Headline check ($Q_6$), which still maintained $>99.9\%$ agreement. The low standard deviation ($\sigma \approx 0.029$) reinforces the finding that high-performing LLM judges exhibit robust verdict stability on qualitative constraints, even when the task involves creative nuance.

\begin{table*}[htbp]
\centering
\small 
\setlength{\tabcolsep}{3.5pt} 
\renewcommand{\arraystretch}{1.3} 

\newcommand{\res}[3]{\makecell{\textbf{#1} \\ (#2, #3)}}
\caption{\textbf{Verdict Stability Analysis (Scenario B: Subjective)}. 
Values formatted as: \textbf{Percentage Agreement, (Gwet's AC1, Fleiss' Kappa)}.
Columns $Q_1$--$Q_7$ represent atomic question stability, while \textit{Rubric $\sigma$} represents the standard deviation of the aggregate adherence score.
\textbf{Key (mapped to Table~\ref{tab:scenario_b_subjective}):} 
$Q_1$: Synthesize history/goals (Cohesive narrative); 
$Q_2$: Re-frame negatives as positive opportunities; 
$Q_3$: Motivational and aspirational tone; 
$Q_4$: Professional yet accessible language; 
$Q_5$: Demonstrates empathy; 
$Q_6$: Punchy/memorable slogan headline; 
$Q_7$: Action-oriented vision statement.}
\label{tab:scenario_b_subjective_verdict}
\resizebox{\textwidth}{!}{%
\begin{tabular}{l c c c c c c c | c}
\toprule
\textbf{Runs ($k$)} & \textbf{$Q_1$} & \textbf{$Q_2$} & \textbf{$Q_3$} & \textbf{$Q_4$} & \textbf{$Q_5$} & \textbf{$Q_6$} & \textbf{$Q_7$} & \textbf{Rubric $\sigma$} \\
\midrule

$k=5$ & 
\res{100}{1.000}{1.000} & \res{100}{1.000}{1.000} & \res{100}{1.000}{1.000} & \res{100}{1.000}{1.000} & 
\res{100}{1.000}{1.000} & \res{99.96}{0.999}{0.989} & \res{100}{1.000}{1.000} & 
0.0133 \\ \addlinespace

$k=10$ & 
\res{100}{1.000}{1.000} & \res{100}{1.000}{1.000} & \res{100}{1.000}{1.000} & \res{100}{1.000}{1.000} & 
\res{100}{1.000}{1.000} & \res{99.98}{1.000}{0.994} & \res{100}{1.000}{1.000} & 
0.0094 \\ \addlinespace

$k=20$ & 
\res{100}{1.000}{1.000} & \res{100}{1.000}{1.000} & \res{100}{1.000}{1.000} & \res{100}{1.000}{1.000} & 
\res{100}{1.000}{1.000} & \res{99.96}{0.999}{0.988} & \res{100}{1.000}{1.000} & 
0.0225 \\ \addlinespace

$k=25$ & 
\res{100}{1.000}{1.000} & \res{100}{1.000}{1.000} & \res{100}{1.000}{1.000} & \res{100}{1.000}{1.000} & 
\res{100}{1.000}{1.000} & \res{99.97}{0.999}{0.991} & \res{100}{1.000}{1.000} & 
0.0158 \\ \addlinespace

$k=30$ & 
\res{100}{1.000}{1.000} & \res{100}{1.000}{1.000} & \res{100}{1.000}{1.000} & \res{100}{1.000}{1.000} & 
\res{100}{1.000}{1.000} & \res{99.91}{0.998}{0.977} & \res{100}{1.000}{1.000} & 
0.0372 \\

\midrule
\textbf{Pooled (90)} & 
\res{100}{1.000}{1.000} & \res{100}{1.000}{1.000} & \res{100}{1.000}{1.000} & \res{100}{1.000}{1.000} & 
\res{100}{1.000}{1.000} & \res{99.95}{0.999}{0.985} & \res{100}{1.000}{1.000} & 
0.029 \\

\bottomrule
\end{tabular}
}
\end{table*}
\end{document}